\documentclass[10pt,conference]{IEEEtran}
\makeatletter
\def\ps@headings{%
\def\@oddhead{\mbox{}\scriptsize\rightmark \hfil \thepage}%
\def\@evenhead{\scriptsize\thepage \hfil \leftmark\mbox{}}%
\def\@oddfoot{}%
\def\@evenfoot{}}
\makeatother
\IEEEoverridecommandlockouts
\usepackage{cite}
\usepackage{amsmath,amssymb,amsfonts}
\usepackage{algorithmic}
\usepackage{graphicx}
\usepackage{textcomp}
\usepackage{comment}
\usepackage[T1]{fontenc}
\usepackage{longtable}
\usepackage{lscape}
\usepackage{caption}
\usepackage{subcaption}
\usepackage{multirow}
\usepackage{graphicx}
\usepackage[table,xcdraw]{xcolor}
\usepackage[normalem]{ulem}

\def\BibTeX{{\rm B\kern-.05em{\sc i\kern-.025em b}\kern-.08em
    T\kern-.1667em\lower.7ex\hbox{E}\kern-.125emX}}

\definecolor{darkgreen}{rgb}{0.29, 0.33, 0.13}
\definecolor{std_red}{rgb}{1, 0, 0}
\definecolor{std_green}{rgb}{0, 1, 0}

\newcommand{\system}{SPON}
\begin{document}
\bstctlcite{IEEEexample:BSTcontrol}

\title{SPON: Enabling Resilient Inter-Ledgers Payments with an Intrusion-Tolerant  Overlay 
}
\author{%
\IEEEauthorblockN{Lucian Trestioreanu}
\IEEEauthorblockA{\textit{Interdisciplinary Centre for}\\\textit{Security, Reliability and Trust}\\\textit{University of Luxembourg}\\
Luxembourg, Luxembourg \\
lucian.trestioreanu@uni.lu}
\and \IEEEauthorblockN{Cristina Nita-Rotaru}
\IEEEauthorblockA{\textit{Khoury College of Computer Sciences} \\
\textit{Northeastern University}\\ Boston, USA \\ c.nitarotaru@northeastern.edu}
\and \IEEEauthorblockN{Aanchal Malhotra}
\IEEEauthorblockA{\textit{Xpring} \\
\textit{Northeastern University}\\ Boston, USA \\ amalhotra@ripple.com}
\and  \IEEEauthorblockN{Radu State}
\IEEEauthorblockA{\textit{Interdisciplinary Centre for}\\\textit{Security, Reliability and Trust}\\\textit{University of Luxembourg}\\
Luxembourg, Luxembourg \\
radu.state@uni.lu}
}
\maketitle


\begin{abstract} 

Payment systems are a critical component of everyday life in our society.
While in many situations payments are still slow, opaque, siloed, expensive or even fail, users expect them to be fast,  transparent, cheap, reliable and global. Recent technologies such as distributed ledgers create opportunities for near-real-time, cheaper and more transparent payments. However, in order to achieve a global payment system, payments should be possible not only within one ledger, but also across different ledgers and geographies.  

In this paper we propose Secure Payments with Overlay Networks (\system), a service that enables global payments across multiple ledgers by combining the transaction exchange provided by the Interledger protocol with an intrusion-tolerant overlay of relay nodes to achieve
(1) improved payment latency,
(2) fault-tolerance to benign failures such as node failures and network partitions, and
(3) resilience to BGP hijacking attacks. 
We discuss the design goals and present an implementation based on the
Interledger protocol and Spines overlay network. We analyze
the resilience of \system~and demonstrate through experimental evaluation
that it is able to improve payment latency, recover from path outages, withstand network partition 
attacks, and disseminate payments fairly across multiple ledgers.
We also show how SPON can be deployed to make the communication between different ledgers  resilient to BGP hijacking attacks.

\end{abstract} 



\section{Introduction}
\label{sec:intro}


Recent technologies such as distributed ledgers (DLT) create opportunities for near-real-time, cheaper, global, and more transparent payments. Examples include Hyperledger Fabric\cite{hyperledger}, R3 Corda\cite{corda}, Quorum\cite{quorum}, Stellar\cite{Stellar}, Overledger\cite{overledger}, OpenChain\cite{openchain}, or private ETH configurations. Central banks are experimenting with DLT: Project Stella between EU and Japan, Jasper and Udin between Canada and Singapore, Project Khokha in South Africa, Emerald at Royal Bank of Scotland, UPI in India, and experimentations at the Central Bank of Brasil are just a few examples. 

Recent developments in protocols allow payments to be initiated on one ledger and cross {\em multiple ledgers} until reaching the final payee, creating a unique opportunity for open and global payment systems. One such solution proposed to perform payments across different ledgers is the Interledger protocol (ILP)~\cite{ILPwhite}. ILP is an application-layer solution and thus, it is not designed to address network level issues such as optimizing for network latency,
or resilience to network level attacks, degradations and failures. When deployed over the Internet, ILP can suffer from service degradations like lossy links, network failures and routing mis-directions of benign or malicious nature. 
Payment systems are critical systems and thus it is desirable they have similar levels of resiliency and security encountered in cyber-physical systems or SCADA~\cite{yadav2020architecture} networks. 

One approach to address the performance, resilience, and security issues is to use an overlay of relay nodes. These relay nodes are not part of
the distributed ledger's nodes and their only goal is to relay communication between ledgers. 
Such an overlay of relays can leverage redundancy in
the  IP network and deploy customized protocols to provide desired security, latency performance, and resilience to failure and attacks.   

Previous work used relays to solve some of these individual problems. For example SABRE \cite{hijackBtc} was proposed to address BGP routing attacks against Bitcoin (BTC); SABRE relies on the BGP path selection to ensure through the placement of a few nodes (<10)  that most BTC nodes will not be partitioned by a BGP hijacking approach. This is achieved by relaying all the traffic through this very small set of relays that must be equipped with sophisticated hardware to sustain the high load of the BTC network. Changes in BGP peering relationships and costs will impact the correct functioning of SABRE. SABRE also relies on the fact that many BTC clients are within a very small number of ASes, and as such, scaling it for inter-ledger communication in order to cover clients spread across many different locations may not be a straightforward task. SABRE does not employ custom protocols to improve performance. Finally, SABRE requires that relay nodes do not get compromised and follow the protocol correctly. Example solutions focused on performance are Falcon~\cite{falconSpringer} and Fibre~\cite{fibre} which both use relays for fast dissemination of BTC blocks, 
and BloXroute \cite{Klarman2018bloXrouteA} which also uses relays for fast  dissemination of blocks for several  ledgers. All of them focus on blocks and not payments, are vulnerable to routing attacks and, as SABRE, assume that the relay nodes are not compromised and follow the protocol correctly. 
%



In this paper we show how a global payment system enabling payments between different ledgers can be designed and deployed over the public Internet using ILP and Spines \cite{spines_intrusion} intrusion-tolerant overlay network. ILP facilitates the interoperability of any payment systems across different ledgers, while Spines serves as secure and trusted transport backbone for ILP communication. We assume that clients conducting payments within the same ledger are handled by internal ledger-specific protocols (e.g. BTC), and  {\em we focus on inter-ledgers communication}. Our Secure Payments with Overlay Networks (\system) system provides (1) improved payment latency between ledgers, (2) fault-tolerance to benign failures such as node failures and network partitions, and (3) resilience to BGP attacks. While intra-ledger protocols typically consider that any ledger node can be compromised (e.g. BTC nodes), previous work using relays to connect ledgers did not assume that relay nodes between ledgers can also be compromised and not forward payments or that the relay network itself can be subject to BGP routing attacks. 

We implemented \system~and investigated how well it achieves its goals. \textit{We consider 3 network topologies}: the first is a synthetic topology allowing to investigate different capabilities of~\system; the second
is based on a real-life deployment \cite{spines_intrusion} with nodes spanning East Asia, North America and Europe and allows  to evaluate~\system's performance in a realistic scenario, and the third 
was used in ~\cite{8613949} 
to show the impact of eclipse attacks conducted by partitioning the network using BGP hijacking and we use it to show
how \system~can be deployed to address such attacks.

We summarize our findings as following:
\begin{itemize}
\item We showed that  \system~improves the payments latency over a \textit{baseline} system not using the overlay.
Benefits become higher as network loss increases, because the customized overlay protocols 
recover the lost packets from nodes closer to the recipient  instead of recovering it from the  sender.
\item We showed that even under extreme scenarios such as a network meltdown \system~was able to continue forwarding payments by rerouting around the failures, while the baseline system could not complete the payments.
\item We use the network topology in~\cite{8613949} 
to show the impact of eclipse attacks conducted by partitioning the network using BGP hijacking, as a demonstrative example on how \system~should be deployed to address such attacks.
\end{itemize}

The structure of the paper is as follows. Section~\ref{sec:bk} discusses challenges for global payment systems and how to overcome them by using overlays. Section~\ref{sec:design} presents our \system~design and implementation, Section~\ref{sec:results} presents our experimental results. 
We place our project in the context of related work in \ref{sec:relwork} and finally, we conclude in Section~\ref{sec:conclusion}. 


\section{Background}
\label{sec:bk}

\subsection{Payments Across Different  Distributed Ledgers with ILP}


One interoperable solution proposed to support payments across different ledgers is the ILP protocol. We consider version 4 of ILP\footnote{https://interledger.org/rfcs/0027-interledger-protocol-4/}, or ILPv4. Its main usage consists  in multi ledger payments, enabled by a set of connectors. To stream payments, the ILP stack provides STREAM, an additional transport protocol which breaks large payments in packets of smaller value.


The ILP ecosystem comprises of multiple software components. \textit{Ledgers} keep records of users accounts and balances, either in fiat or crypto-currencies. \textit{Connectors} are the transaction intermediaries and  hold multiple wallets on different ledgers, such that they can perform currency exchange, and forward packets on behalf of their customers, while receiving a fee. Finally,  \textit{Applications} run by end-users to perform transactions; examples include \textit{Moneyd}, or \textit{Switch} by Kava Labs.

\begin{figure}[ht]
\begin{center}
    \includegraphics[width=0.45\textwidth]{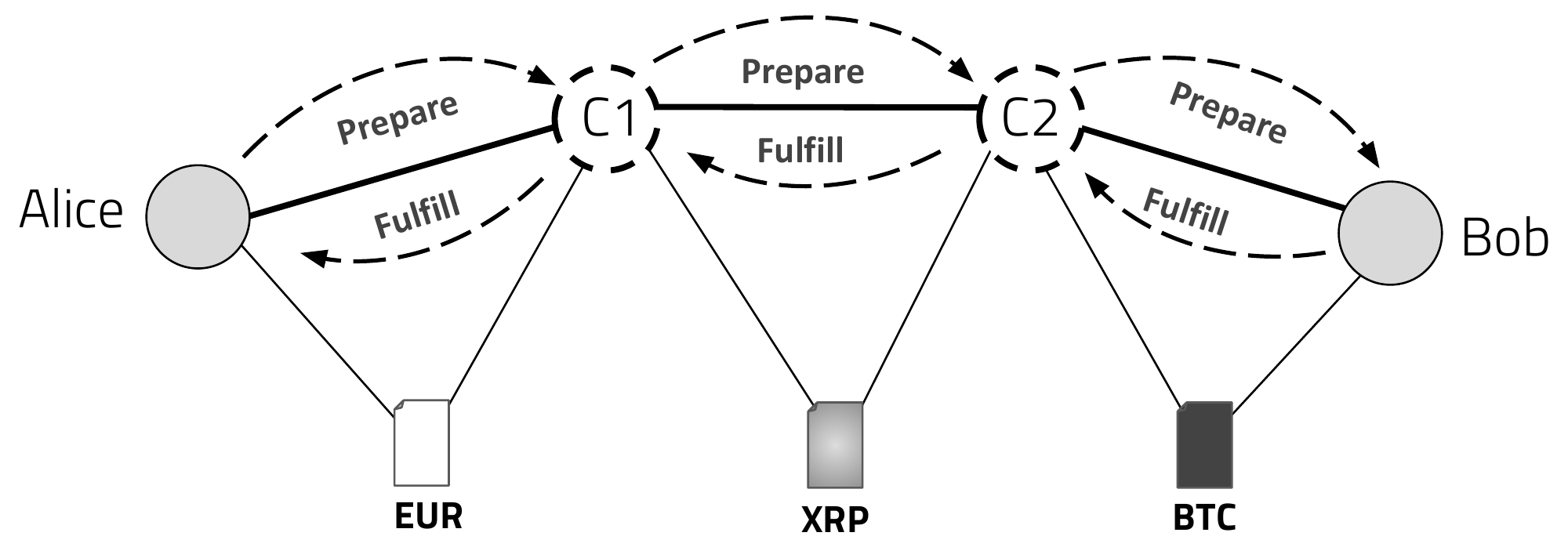}
    \caption{Payment with ILP. C1 and C2 are ILP connector nodes.}
    \label{fig:ILP_basics}
\end{center}
\end{figure}

Figure~\ref{fig:ILP_basics} shows how ILP facilitates payments. 
Consider customers \textit{Alice} and \textit{Bob}, where \textit{Alice} has an account in Euro and wants to pay \textit{Bob}, who has an account in BTC. Connector \textit{C1} has an account in Euro, and an account in XRP, while Connector \textit{C2} has an account in XRP and an account in BTC. C1 and C2 are peered together, i.e. they negotiated also a business relationship. ILP allows Alice to create a payment request in Bob's favor, which will travel from her to C1, C2 and then to Bob. Upon receiving the payment, Bob will send back on the same path a receipt, which will finally reach Alice. The receipt assures all parties that the payment was successful and they settle their balances. As it travels between connectors, the value changed wallets and currencies. 


%

\subsection{Limitations of ILP Payment Systems over the Internet}

\begin{figure}[ht]
\begin{center}
    \includegraphics[width=0.485\textwidth]{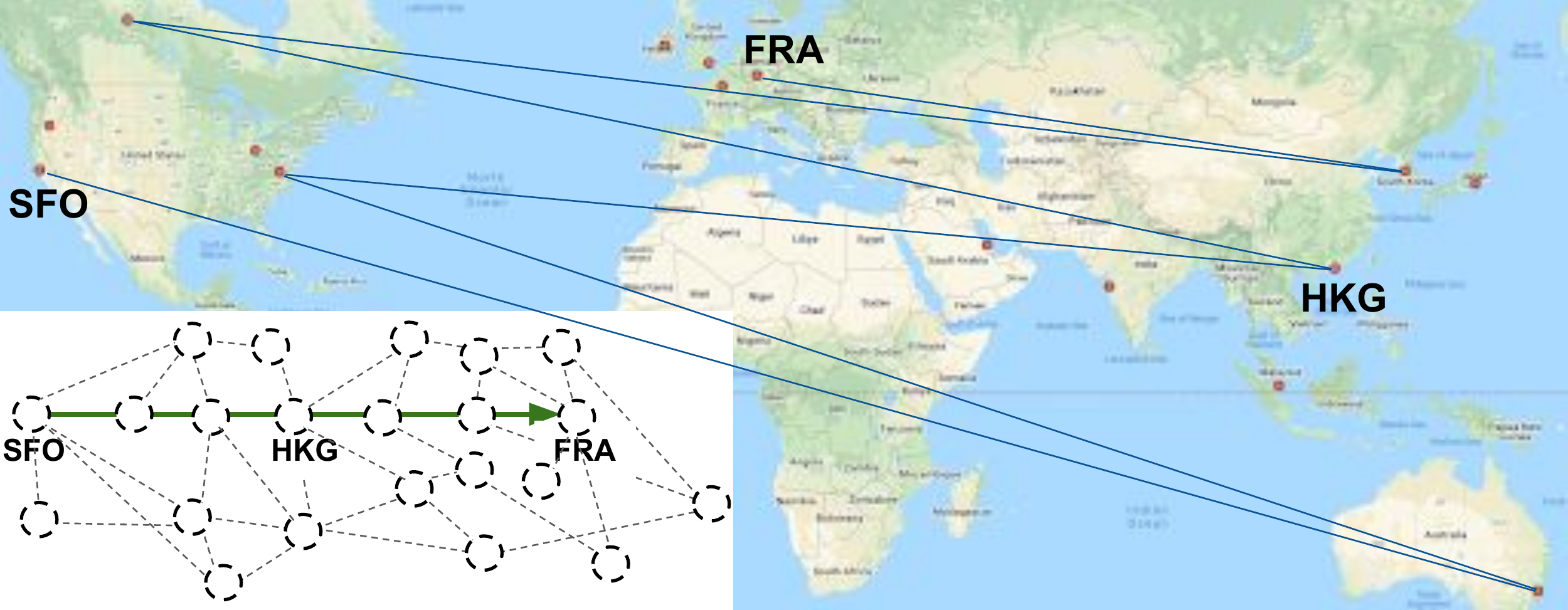}
    \caption{Example ILP payment routing (lower left thumbnail) and actual geographical location of corresponding ILP nodes.}
    \label{fig:long_path}
\end{center}
\end{figure}

To facilitate the discussion about some of the limitations of current payment systems designs we present an example in Figure~\ref{fig:long_path}. The lower left thumbnail shows a possible example of an ILP network, where the nodes are ILP connectors. As ILP nodes may freely form links on the ILP network, according to reasons like regulatory, legal, business and trust relationships, the network is not constructed based on latency or attack resilience criteria. So, according to current ILP payment routing algorithm, a payment from San Francisco (SFO) to Frankfurt (FRA) could be routed along the green path also including Hong Kong (HKG). The physical locations of these ILP nodes along the payment path highlighted in green could be spread all around the world, resulting in high end-to-end latency and increased vulnerability of the payment system to lossy data paths, faults and attacks.

%
In this paper we focus on network level limitations of ILP payment systems. We identify three such limitations: (1) resilience to  lossy paths, (2) resilience to network faults and partitions, (3) resilience to DoS such as route hijacking. 

{\em Lossy paths} can be problematic especially in case of streaming payments, in which one single payment can be spawned over multiple smaller payments. This is encountered in  pay-as-you go for torrent like distribution services \cite{ilptorrent}, which can not afford packet losses even if the per packet level payment amount is tiny. Many underbanked communities \cite{ruralFintech} experience the downsides of digital, financial divides
and even in developed economies some rural communities have to face mediocre Internet connectivity. 


{\em Path failures and network partitions.}
Network resilience is an important factor to consider since  network enabled systems  can be partitioned by intentional actions (censorship) or non-intentional (faults) accidents.  The consequences for both are the same: outage, delays and  degraded performance which impact the availability of the service. Payment systems should be  capable to rapidly detect failures and react accordingly.  

{\em BGP hijacking attacks.}
        BGP routing attacks against ILP could have a serious impact such as:
        partition the payment network and create a situation similar to a DoS, which can result in revenue loss for ILP nodes and their customers (open attack), 
delay all/chosen packets, while attacker’s packets would be forwarded at normal rate (covert attack),
 hairpin drop packets from/to a certain ILP node/endpoint (covert attack), or
at will, attacker can be the only one able to send/receive ILP transactions in/from both partitions.
 The attacker can also divert, store, map and analyse the traffic: get geo-location information of ILP providers/customers, gather/infer information about payments volumes per ILP node (average value carried by an ILPv4 packet at the attack moment is $x$ XRP). 

\section{\system~Design and Implementation}
\label{sec:design}

In this section we describe \system,~our proposal for resilient global payment systems over Internet.
We first describe the design goals for our system, then describe the attacker model, and give a description of the design and implementation.

\subsection{Design Goals and High-level Approach}

Our main goal is to design a global payment system that supports payments across different ledgers \textit{while achieving:} improved performance (latency), improved service availability (fault-tolerance), and security guarantees, including resilience to routing attacks. We assume clients conducting payments within the same ledger are handled by ledger-specific protocols. While these internal protocols can also benefit from additional improvements, {\em our focus is on connecting different ledgers and not on services within a ledger}. We use ILP to facilitate the exchanges across different currencies and ledgers. However, ILP is not meant to optimize network communication and address fault-tolerance to network failures or BGP attacks. With our goals in mind, we would like our service connecting multiple ledgers to have the following properties:

\begin{enumerate}
  \item[\textbf{G1}]  {Improved payment latency}: Our design should leverage the redundancy in the underlying IP network to take advantage of links offering better connectivity, by using customized routing protocols. 
  \item[\textbf{G2}]  {Resilience to lossy paths}: Our design should be resilient to lossy communication links across ledgers and as such improve the client network's resilience to lossy links.
  \item[\textbf{G3}]  {Resilience to path failure and node crashes}: The design should increase payment service availability by increasing data flow availability through providing a system resilient to network path failures and relay node crashes.
  \item[\textbf{G4}]  {Resilience to BGP routing attacks}: Our design, also deployment dependent,  should provide resilience to routing attacks like \textit{Coremelt} and \textit{Crossfire}~\cite{coremelt,crossfire}.
  \end{enumerate}

  {\bf Approach.} These goals can be achieved by changing an existing payment-exchange protocol like ILP to add the desired performance, fault-tolerance, and attack resilience. However, we argue that a separation between the payment-exchange and the communication functionalities provides more flexibility in ILP node placement and modularized development. For example, the ledger pre-post processing functionality is better placed closer to the ledger; also, because they manipulate user value and data, the placement of ILP nodes in different geographical areas may involve different legal restrictions, licensing, regulations. A compromised ILP node is more dangerous than a compromised overlay node performing a simple forwarding because the forwarding nodes do not need visibility into the payments to perform network-level forwarding. Thus, our approach is to separate ledger processing from the forwarding functionality, to maximize performance and resilience to attacks, while accommodating legal restrictions. The data forwarding layer can be an overlay of {\em relay nodes} that implement customized routing algorithms for better latency, routing around failures and with BGP attack resilience. The ILP payment exchange connectors use the overlay of relays to communicate with each other. 

\begin{figure}[ht]
\begin{center}
    \includegraphics[width=0.485\textwidth]{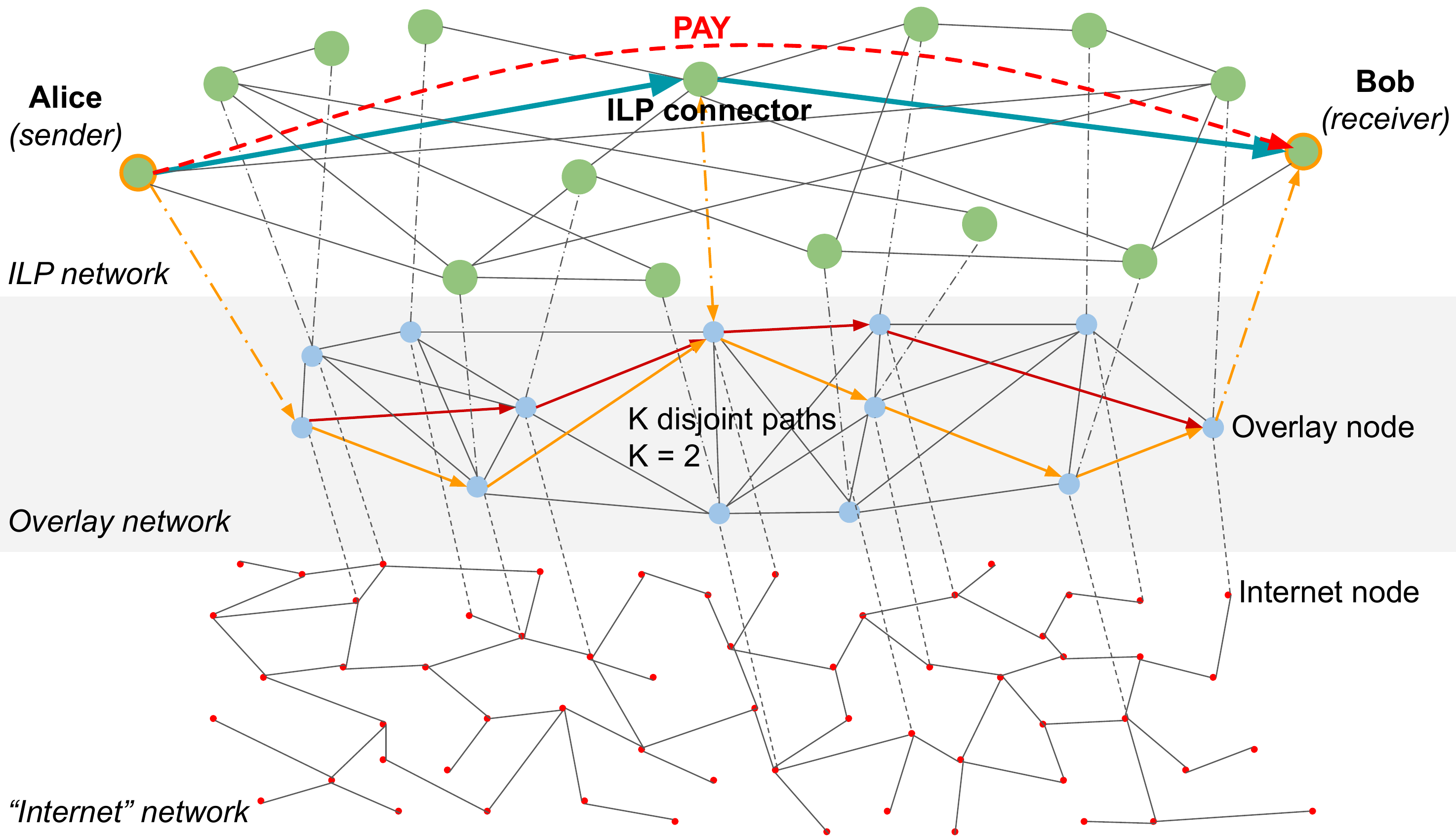}
    \caption{Communication mapping for Ledgers, overlay, Internet.}
    \label{fig:ILP_Spines_IP}
\end{center}
\end{figure}

  Figure \ref{fig:ILP_Spines_IP} shows how communication flows between ILP nodes Alice and Bob, through ILP and the overlay of relay nodes (Alice and Bob are not end-users but full ILP nodes): Each ILP node is connected to at least one overlay relay node. Each overlay relay node is connected to multiple Internet Service Providers (ISP) / Internet Exchange Points (IXP) / Autonomous Systems (AS). At ILP level, a payment originated from Alice for Bob, is routed through the "ILP connector" in the middle. However at data packet level, the 2 hops (Alice <-> Connector and Connector <-> Bob, are routed through redundant paths on the overlay network (thick arrows on the middle layer of Figure \ref{fig:ILP_Spines_IP}). Further, each overlay link benefits from disjoint, redundant paths at Internet level below.
  
 \textbf{Need for intrusion-tolerant overlays.} Overlay networks can improve latency because they can reduce re-transmissions~\cite{claudiu_thesis, 7980115} and can provide resilience to benign faults by routing around them. However, the introduction of the overlay of relay nodes in the system design changes the trust model. First, the overlay itself is susceptible to compromises since a software node is easier to compromise than a hardware router. Compromised overlay nodes can significantly impact the system performance as a whole, or target specific connectors or ledgers and discriminate against some clients conducting payments. Second, the nature of the overlay requires different payment streams to share the same logical structure which can allow some clients to create denial of service against competitor clients conducting payments through the same link(s) on the relay network. Such overlays need to be centrally managed  to prevent topology related attack.  We set the following goals for our overlay of relays:

  \begin{enumerate}
  \item[\textbf{O1}]  {Resilience to attacks from compromised forwarding relays}: We want to prevent compromised relay nodes from being able to divert or stop traffic. 
  \item[\textbf{O2}]  {Resilience to denial-of-service from malicious clients}: In the presence of the overlay, payment flows from different competitor clients can potentially compete to each other at networking level to the point where one can generate a targeted denial of service for the other by saturating the link(s). We would like all payment flows to be treated fairly by the relay nodes, i.e. all payment streams receive the same share of available network bandwidth.
    \end{enumerate}

 \subsection{Threat Model}

We assume that the overlay of relay nodes is centrally managed and communication between relay nodes is authenticated using Public Key Infrastructure (PKI), where the system administrator and each overlay node has a public/private key pair and knows all the other public keys. The overlay topology is known by all of the overlay nodes, and changes can be made only by the system administrator.

We also assume that overlay relay nodes can be compromised. A compromised node can exhibit Byzantine behavior such as arbitrary dropping, delaying, or incorrect forwarding of packets. We assume that overlay nodes have sufficient computational resources to keep up with processing incoming messages, but bounded buffers for message storing.

We do not assume a specific bound on the number of compromised relays in the overlay network. Instead we assume that the adversary cannot partition the sender from the receiver, i.e. there is a path from the sender to receiver where all relays are not controlled by the adversary.

We assume attackers have large amounts of network bandwidth, memory and computation, such as those required by large-scale DDoS attacks as those in \cite{coremelt,crossfire}.

%
%
%

\subsection{\system~Design and Implementation}

We implemented \system~using ILP and the Spines overlay. Below, we first give a description of aspects of ILP and Spines relevant to our design, then describe our system, \system. 
%
%

The ILP environment consists of a  stack of protocols:
\begin{itemize}
    \item \textit{Bilateral Transfer Protocol} (BTP), responsible of establishing a link between two peers.
    \item ILP itself, ensuring the value transfer across ledgers. The ILP packet offers a \textit{data} field of size 32k, where different information and sub-protocols can be encapsulated.
    \item  \textit{Streaming Transport for the Realtime Exchange of Assets and Messages} protocol (STREAM), implementing the concept of streaming value (money) and data over ILP (encapsulated in ILP packets). This concept offers a series of advantages over sending a transaction in full.
    \item \textit{Simple Payment Setup Protocol} (SPSP) ensuring the exchange of credentials required to establish a STREAM payment, which for specific reasons works over HTTP.
\end{itemize}

Spines is an open source overlay network~\cite{7980115,spinesorg} that
 provides availability, resiliency, and timed-delivery, achieved by making use of multi-homing at multiple ISPs and deploying the nodes in strategically located datacenters (connectivity). The nodes are centrally managed
 and resilient overlay routing such as multiple disjoint paths and flooding~\cite{spines_intrusion}-p6 are used to ensure resilience to
 forwarding attacks. Buffer management like round robin is used to ensure that each node evenly processes
 packets per sender in case of priority sending, or per flow (sender-receiver pairs) in case of reliable sending.
 

\begin{figure}[ht]
\begin{center}
    \includegraphics[width=0.485\textwidth]{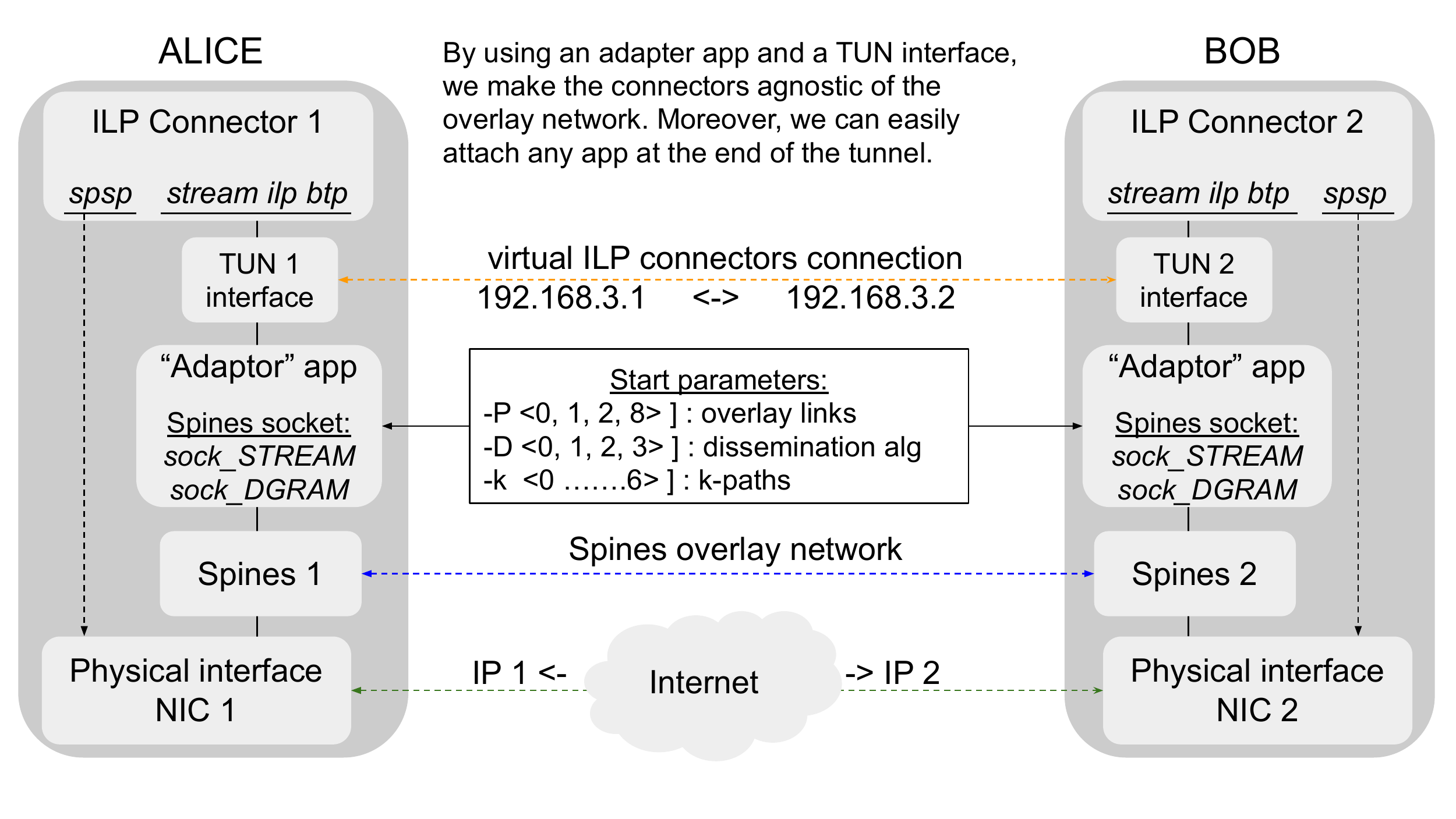}
    \caption{\system~Architecture.}
    \label{fig:system}
\end{center}
\end{figure}

We show the architecture of \system~in Figure \ref{fig:system}. There are 3 network layers: the base internet layer, the Spines overlay, and the ILP network, each featuring their own addressing schemes and protocols. Each ILP node connects to a Spines node using the stack illustrated in Figure~\ref{fig:system}. The connector applications connect through a tunnel, agnostic of the overlay below. An \textit{adapter} application makes the connection to the \textit{spines\_socket} exposed by the Spines node, and sends it the different parameters to use in order to forward data. We use the \textit{Priority Messaging (PRI)} and \textit{Reliable Messaging (REL)} communication services, shown and explained in Table~\ref{tab:spinesParamDetails}. 

One advantage of \system~is that the service can be selected per ILP packet, because Spines provides its reliable or priority services on a per packet basis. Our design exposes this functionality to ILP payments and other ILP tools such as ILP-ping. As such, for example, the risk of \textit{fulfillment failure} specific to ILP, could now be alleviated by prioritizing the \textit{fulfilling} over the \textit{prepare} packets\footnote{https://interledger.org/rfcs/0018-connector-risk-mitigations/}. As needed, any ILP related flow can be \textit{prioritized} or sent \textit{reliably}, for example routing updates or SPSP data could use the reliable protocol.

Because the connectors are agnostic of the overlay below, our design also allows for a \textit{partial deployment}, where some connectors choose to join the network and others do not. This involves the existence of some \textit{bridge} connectors, having connections both outside and inside \system.

\begin{table}[ht]
\centering
\caption{SPON services (via Spines).} 
\label{tab:spinesParamDetails}
\resizebox{0.485\textwidth}{!}{%

\begin{tabular}{rll}
\hline
\rowcolor[HTML]{EFEFEF} 
Service    & Details                                          \\ \hline
\begin{tabular}[c]{@{}r@{}}\textbf{PRIORITY (PRI)}\end{tabular} &
  \begin{tabular}[c]{@{}l@{}}Source-based routing with timeliness guarantees, \\
  i.e. packets are sent based on their priority,\\ each node forwards packets fairly across all sources. 
  \end{tabular} \\ \cline{3-3} 
\begin{tabular}[c]{@{}r@{}}\textbf{RELIABLE (REL)}\end{tabular} &
  \begin{tabular}[c]{@{}l@{}}Source-based routing with reliability guarantees,\\
  i.e. packets are sent with end-to-end reliably,  \\ each node forwards packets fairly across all sender-receiver pairs.
\end{tabular}
\end{tabular}%
}
\end{table}

\section{Experimental Results}
\label{sec:results}

In this section we describe the evaluation of \system. We seek to answer the following questions:

\begin{enumerate}
  \item[\textbf{Q1}] What are the latency improvements of \system~when  compared with an approach that does not use relays?
  \item[\textbf{Q2}] How does \system~react to more severe network events such as network meltdowns?
  \item[\textbf{Q3}] How does \system~handle denial of  service attacks where some clients try to overload the links with payments?
  \item[\textbf{Q4}] How does \system~react to severe network events such as route misdirections and BGP hijacking attacks?
  \end{enumerate}

\subsection{Methodology}

 We conduct our experiments using Mininet  to better control the network topology, links and their properties. We used the "reference" ILP connector\footnote{https://github.com/interledgerjs/ilp-connector} and a private XRP ledger.

\begin{figure}[ht]
\begin{center}
    \includegraphics[width=0.485\textwidth]{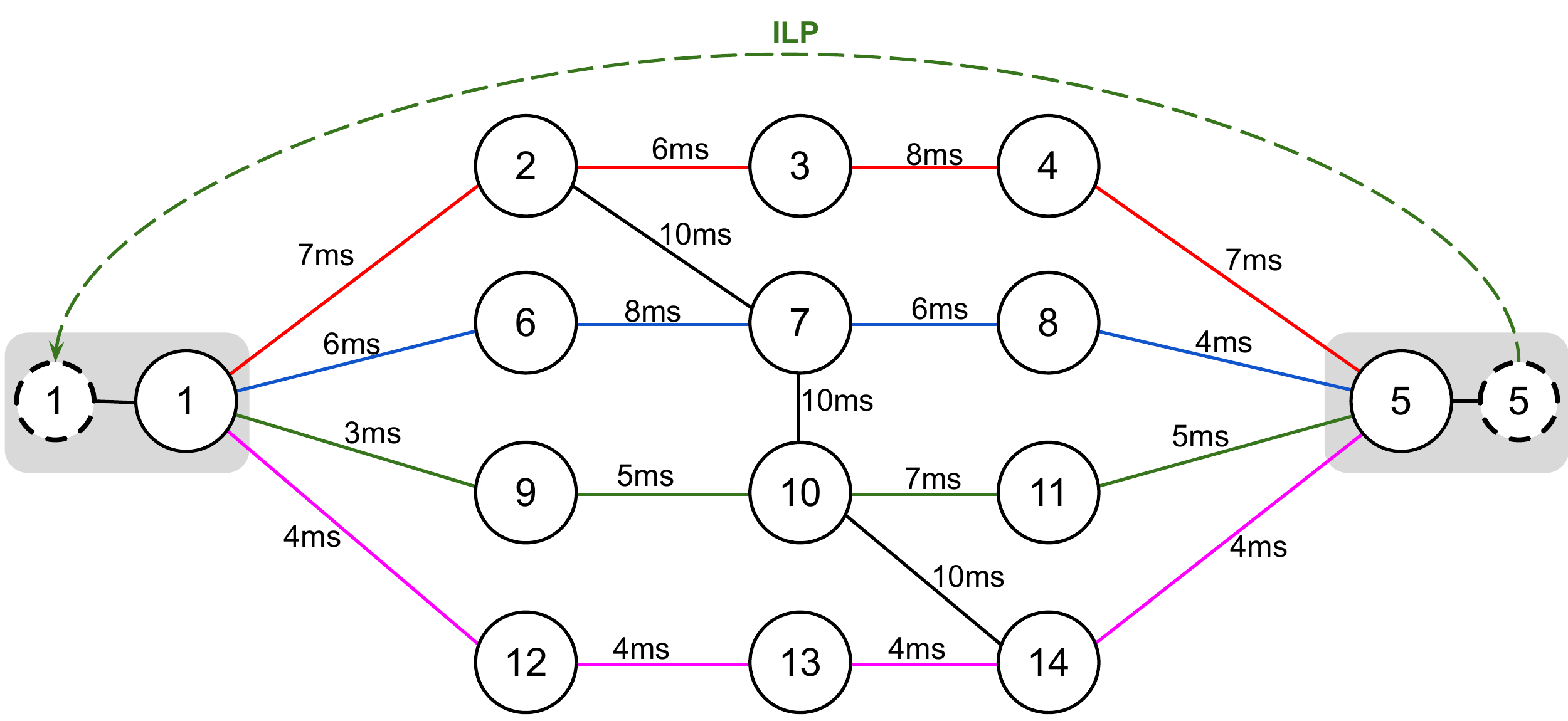}
    \caption{Chain Topology.}
    \label{fig:chain_topology}
\end{center}
\end{figure}

\begin{figure}[ht]
\begin{center}
    \includegraphics[width=0.485\textwidth]{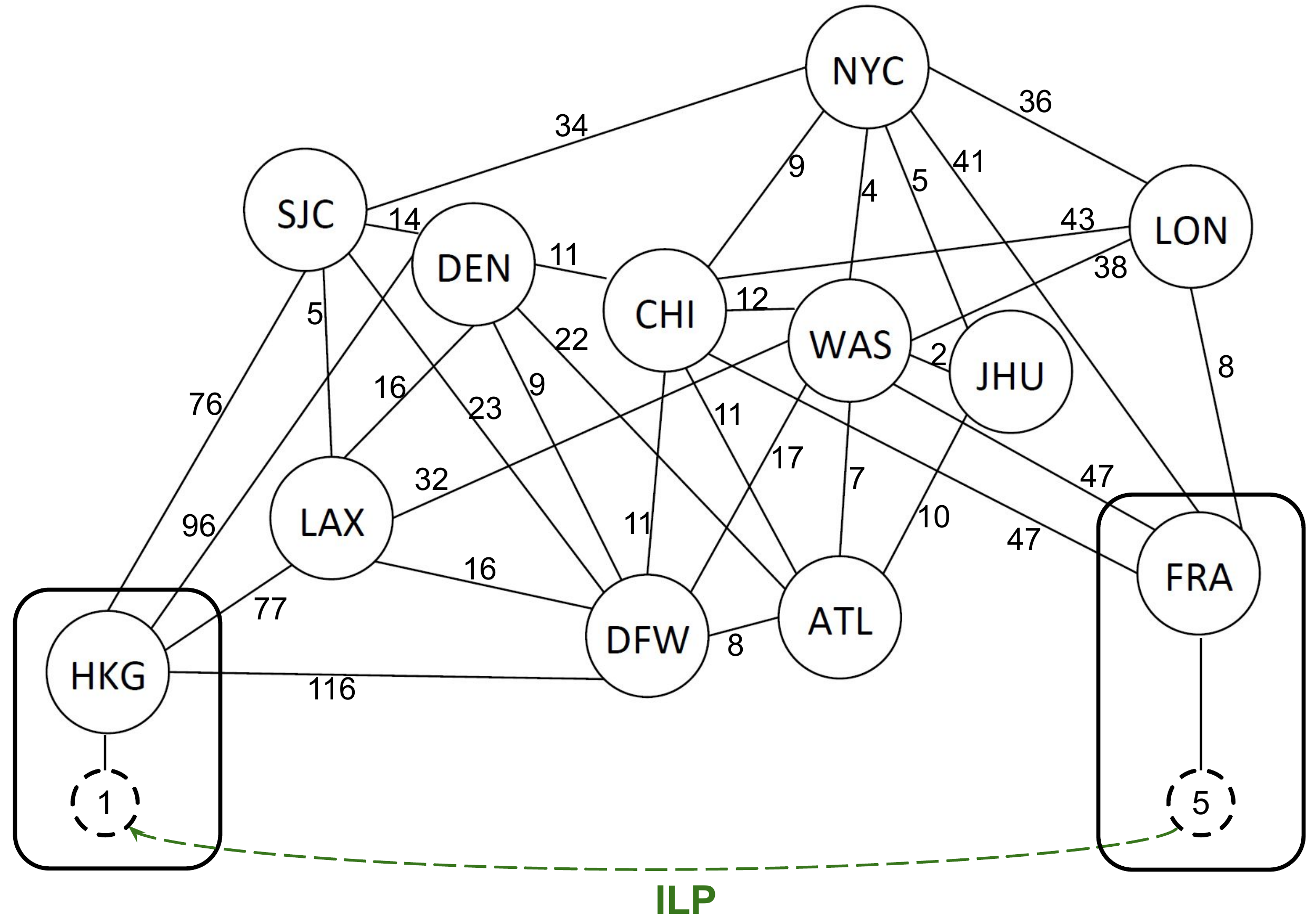}
    \caption{Global Topology.}
    \label{fig:global}
\end{center}
\end{figure}

\emph{Topologies.}
We used 3 topologies for our evaluations, and a fourth to demonstrate BGP resilience. The first, referred as \textit{Chain Topology} (Figure~\ref{fig:chain_topology}) is a demonstrative topology allowing to investigate different path capabilities of our overlay.
The second, referred as \textit{Global Topology} (Figure~\ref{fig:global}) is a real-life topology spanning the Internet and obtained from \cite{spines_dissemination_graphs} which allows to demonstrate the performance and resilience of \system~in a more realistic scenario. Link latencies were obtained from specialized websites\footnote{https://ipnetwork.windstream.net/, https://wondernetwork.com/pings}. Third setting, shown in Figure~\ref{fig:Fairness_topo} helps answer Q3, while Q4 is discussed using Figure~\ref{fig:bgp_mitigation}.

\emph{Systems.} We compare the following configurations:
\begin{itemize}
\item Baseline: payments are sent via ILP nodes  without \system.
\item Priority (PRI): payments use \system~configured with source-based routing and timeliness delivery~\cite{spines_intrusion}.
\item Reliable (REL): payments use \system~configured with source-based routing and reliable delivery~\cite{spines_intrusion}. 
\end{itemize} 
For both Priority and Reliable settings, we evaluated Flooding (FLD) and k-path as communication mechanisms. \textbf{Q1} and \textbf{Q2} are answered by comparing the \textit{Baseline} with SPON's behavior in \textit{PRI} and \textit{REL} mode.

\emph{Metrics.} We use
  \textit{Round Trip Time on ILP} $(RTT_{ILP})$ reported by the ILP Ping tool \footnote{https://github.com/martinlowinski/ilp-ping} to evaluate the communication between ledgers via \system.
For larger payments which are broken into a number of ILP packets and sent via \textit{STREAM},
we use \textit{Payment Latency} as the total time to complete a payment.



\subsection{Performance}

\subsubsection{\textit{Chain topology}} 
As illustrated in Figure~\ref{fig:chain_topology}, we use two ILP nodes ({\em 5} and {\em 1}) acting as sender and receiver, to send 100 ILP ping packets at a rate of 1 packet/s, using the ILP-PING tool. The baseline $(RTT_{ILP})$ is 32ms and equivalates the two \textit{connectors} paired directly on the fastest path from the figure.

\begin{figure*}[t!]
    \centering
    \begin{subfigure}[t]{0.24\textwidth}
        \centering
        \includegraphics[height=1.15in]{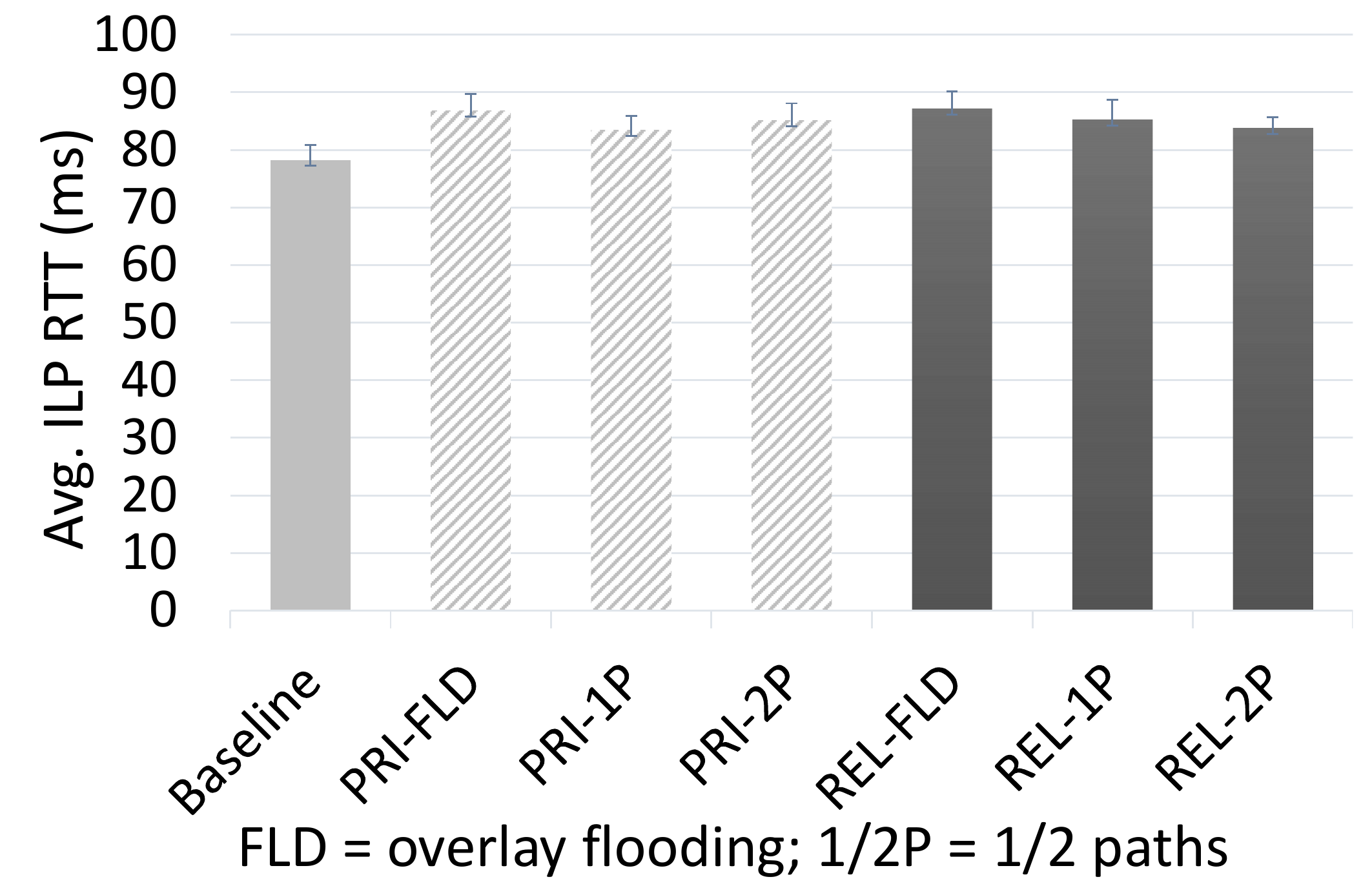}
        \caption{Loss 0\%}
        \label{fig:Experiment_Basic_Latency_Results}
    \end{subfigure}%
    \begin{subfigure}[t]{0.24\textwidth}
        \centering
        \includegraphics[height=1.15in]{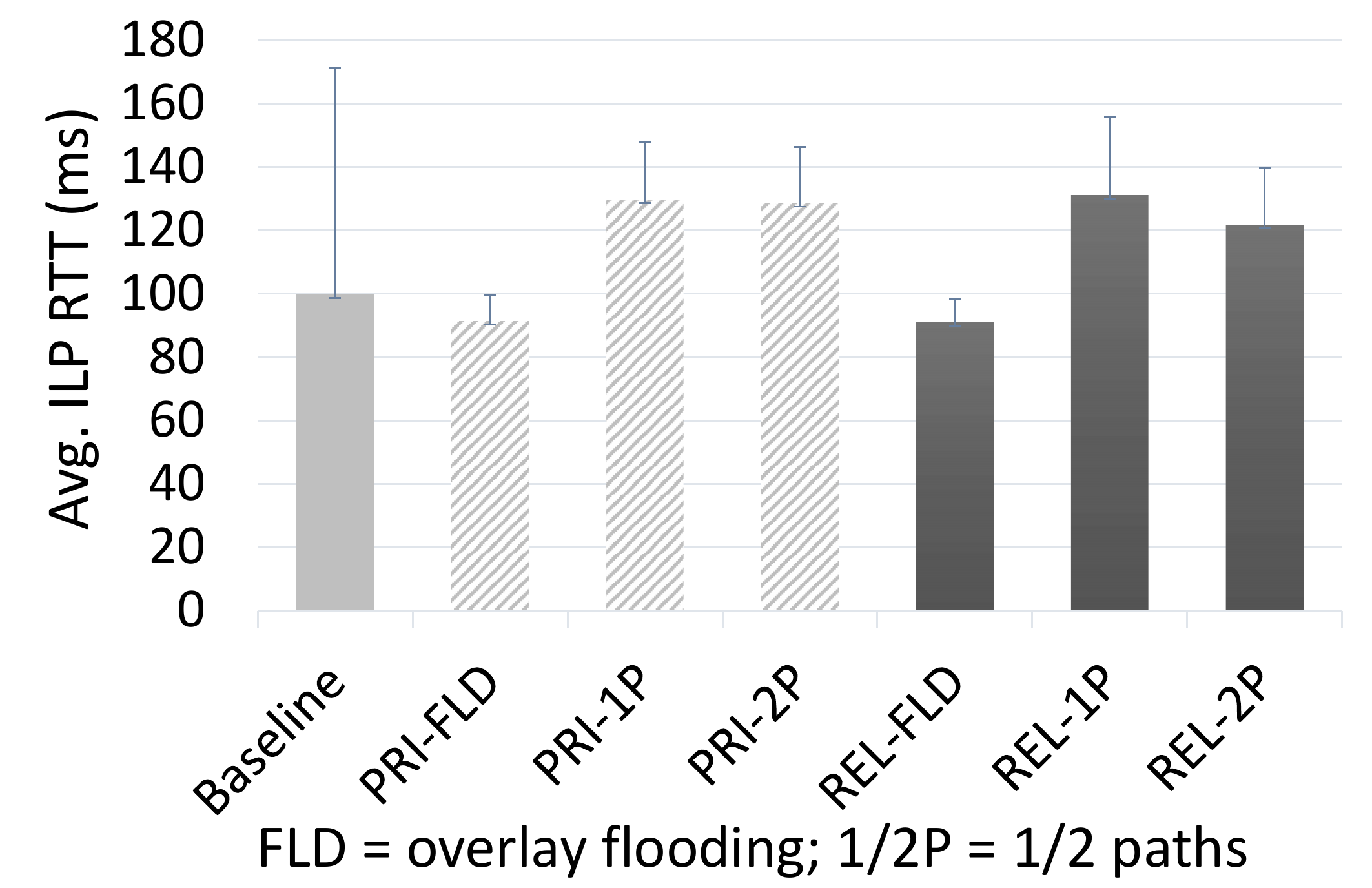}
        \caption{Loss 2\%}
        \label{fig:Experiment_Basic_Latency_Results_2pct}
    \end{subfigure}%
    \begin{subfigure}[t]{0.24\textwidth}
        \centering
        \includegraphics[height=1.15in]{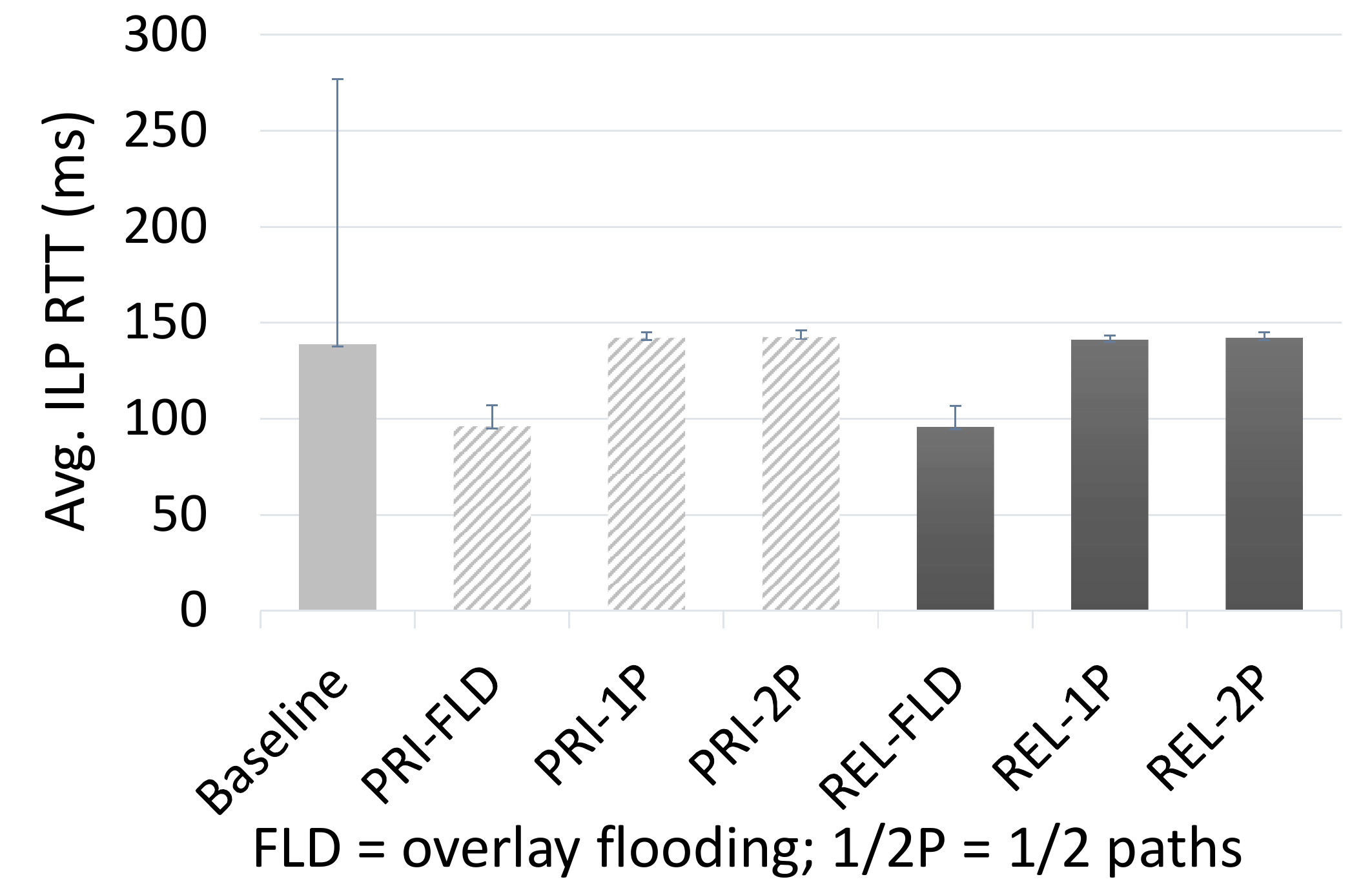}
        \caption{Loss 5\%}
        \label{fig:Experiment_Basic_Latency_Results_5pct}
    \end{subfigure}%
    \begin{subfigure}[t]{0.24\textwidth}
        \centering
        \includegraphics[height=1.2in]{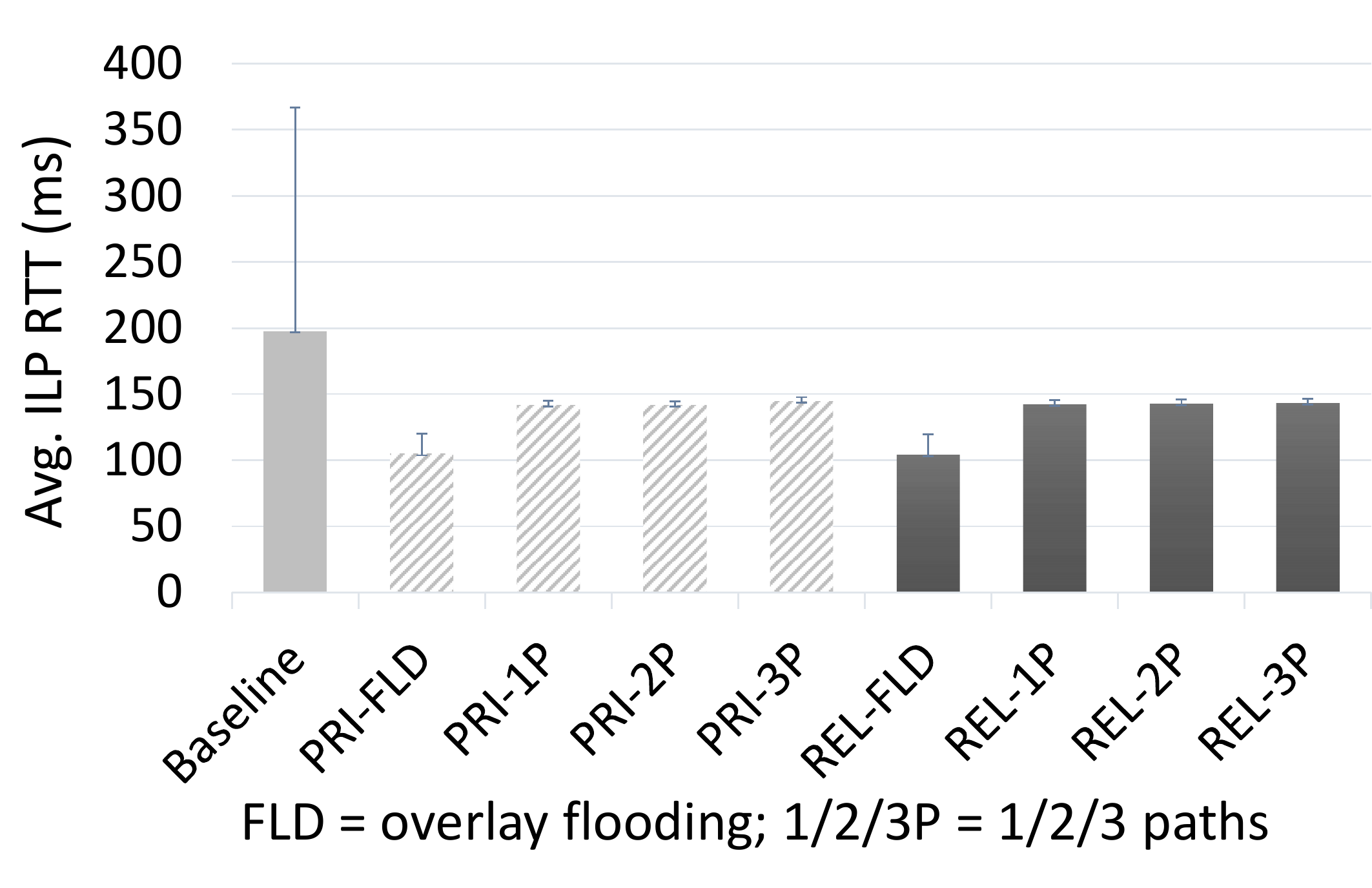}
        \caption{Loss 10\%}
        \label{fig:Experiment_Basic_Latency_Results_10pct}
    \end{subfigure}
    \caption{Average ILP ping RTT on the Chain topology in a network loss scenario, Priority (PRI) or Reliable (REL) messaging.}
\end{figure*}


{\em ILP RTT.} To evaluate latency under loss, we introduce variable loss of 2, 5, 10\% on link S12-S13, chosen because it's on the fastest topology path, so it has high chances to have a visible impact on results, illustrated in Figures~\ref{fig:Experiment_Basic_Latency_Results_2pct},~\ref{fig:Experiment_Basic_Latency_Results_5pct},~\ref{fig:Experiment_Basic_Latency_Results_10pct}. Solid grey bars represent baseline averages, grey striped bars represent \textit{Priority} messaging with flooding (FLD), 1 or 2 paths~\cite{spines_intrusion}-p6, and dark grey bars represent \textit{Reliable} messaging with FLD, 1 or 2 paths. While not shown experimentally, we appreciate that introducing loss on slower paths (9-10, 6-7, 2-3) would advantage SPON by enabling it to use the fastest path at full capability. We isolate Spines' processing overhead by setting loss to 0; as shown in Figure~\ref{fig:Experiment_Basic_Latency_Results}, \system~does fare a little bit worse than the baseline (5\% or 6s in our case). This overhead however is small and does not prevent \system~from performing better than the baseline in realistic situations with loss: at 2\% loss, Figure~\ref{fig:Experiment_Basic_Latency_Results_2pct} shows that SPON already offers an advantage of 10\% latency over the baseline when working in FLD mode. As loss increases, \system's advantage increases, and at 5\% loss the gain over same baseline is 33\%, as depicted in Figure \ref{fig:Experiment_Basic_Latency_Results_5pct}. 
The error bars also point that the service is more stable under loss, if using SPON.

\begin{figure*}[t!]
    \centering
    \begin{subfigure}[t]{0.25\textwidth}
        \centering
        \includegraphics[height=1.15in]{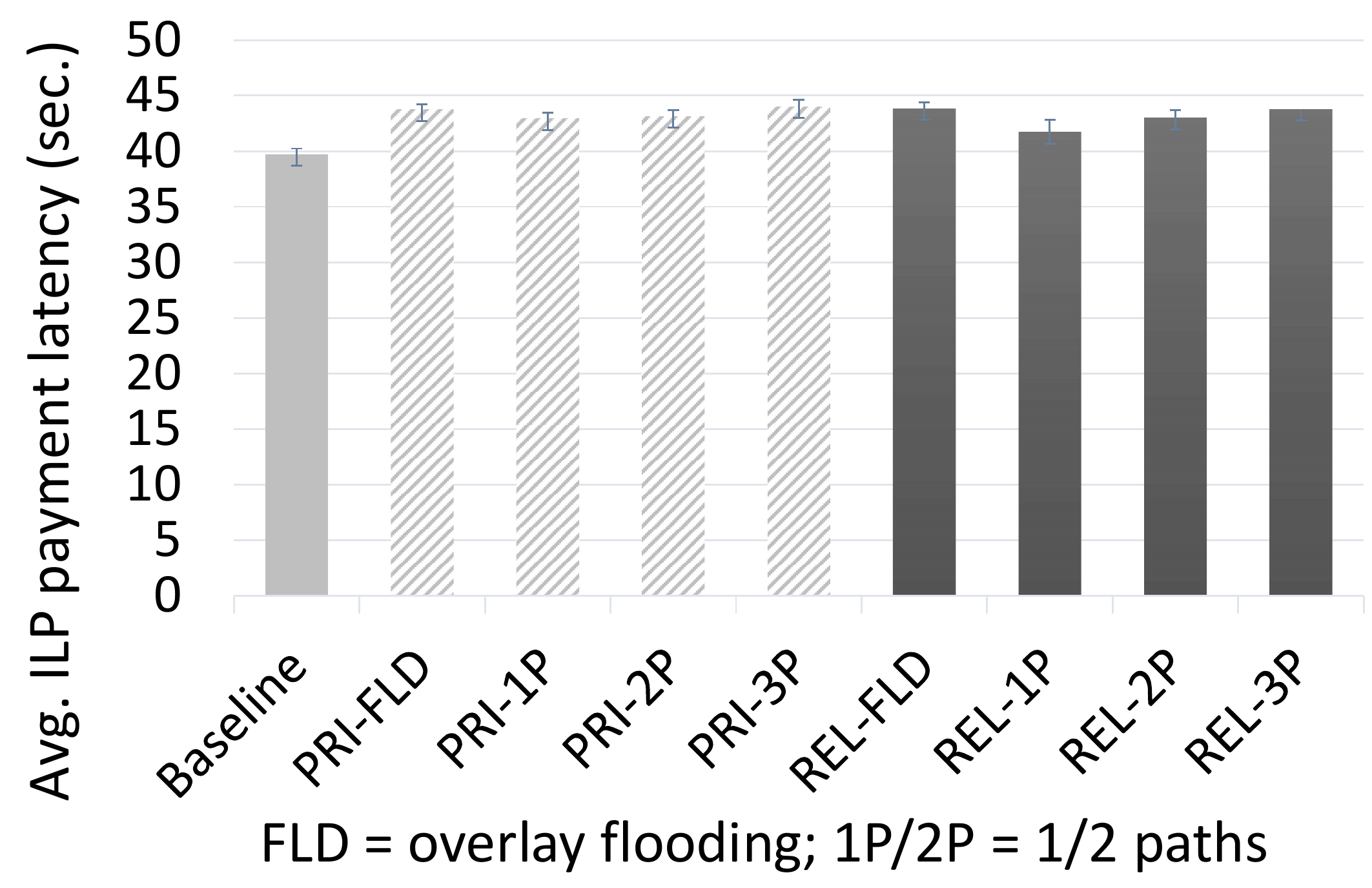}
        \caption{Loss 0\%}
        \label{fig:Exp3a_ILP_STREAM_Latency_0pct}
    \end{subfigure}%
    \begin{subfigure}[t]{0.25\textwidth}
        \centering
        \includegraphics[height=1.15in]{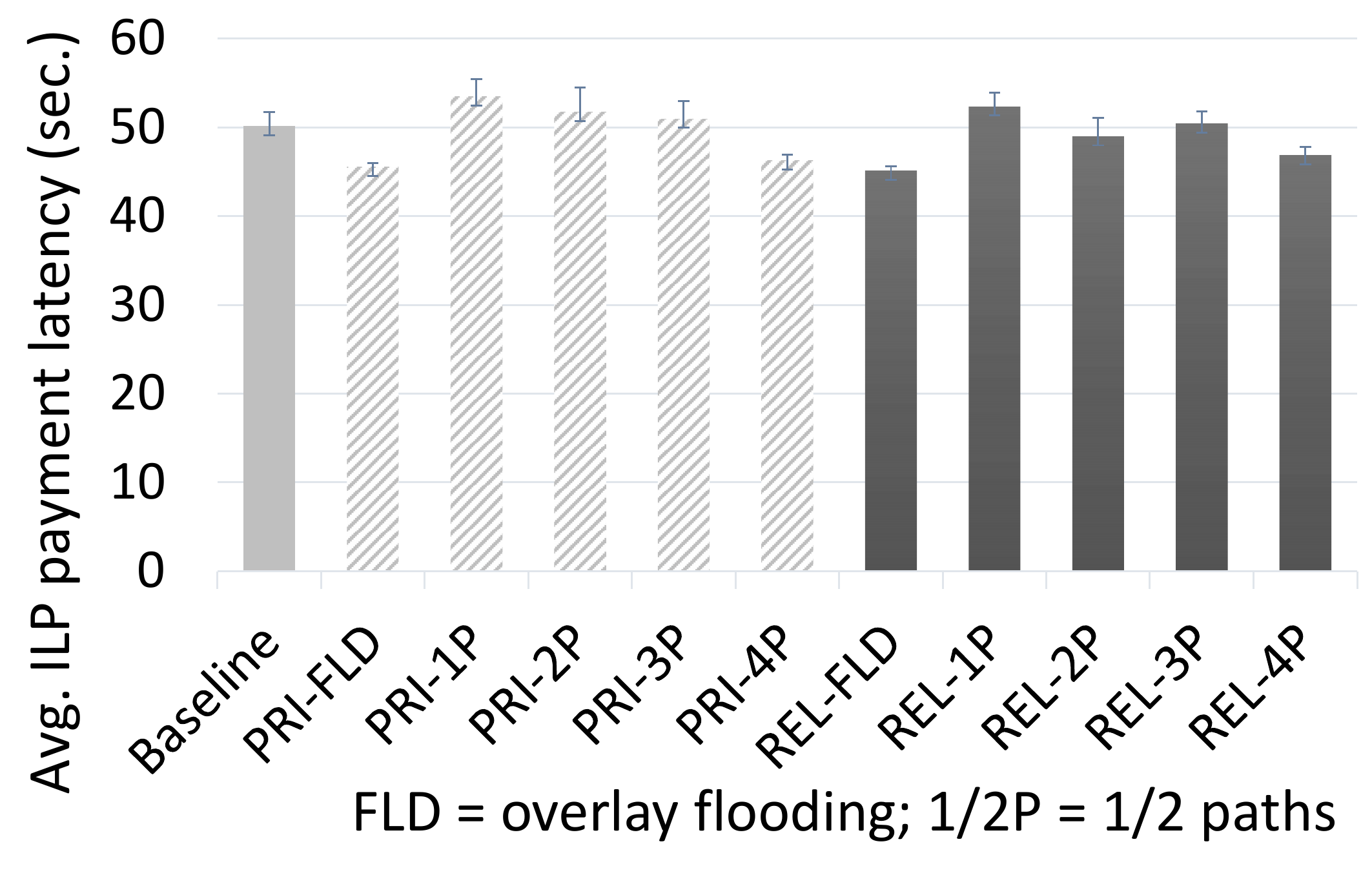}
        \caption{Loss 2\%}
        \label{fig:Exp3a_ILP_STREAM_Latency_2pct}
    \end{subfigure}%
    \begin{subfigure}[t]{0.23\textwidth}
        \centering
        \includegraphics[height=1.13in]{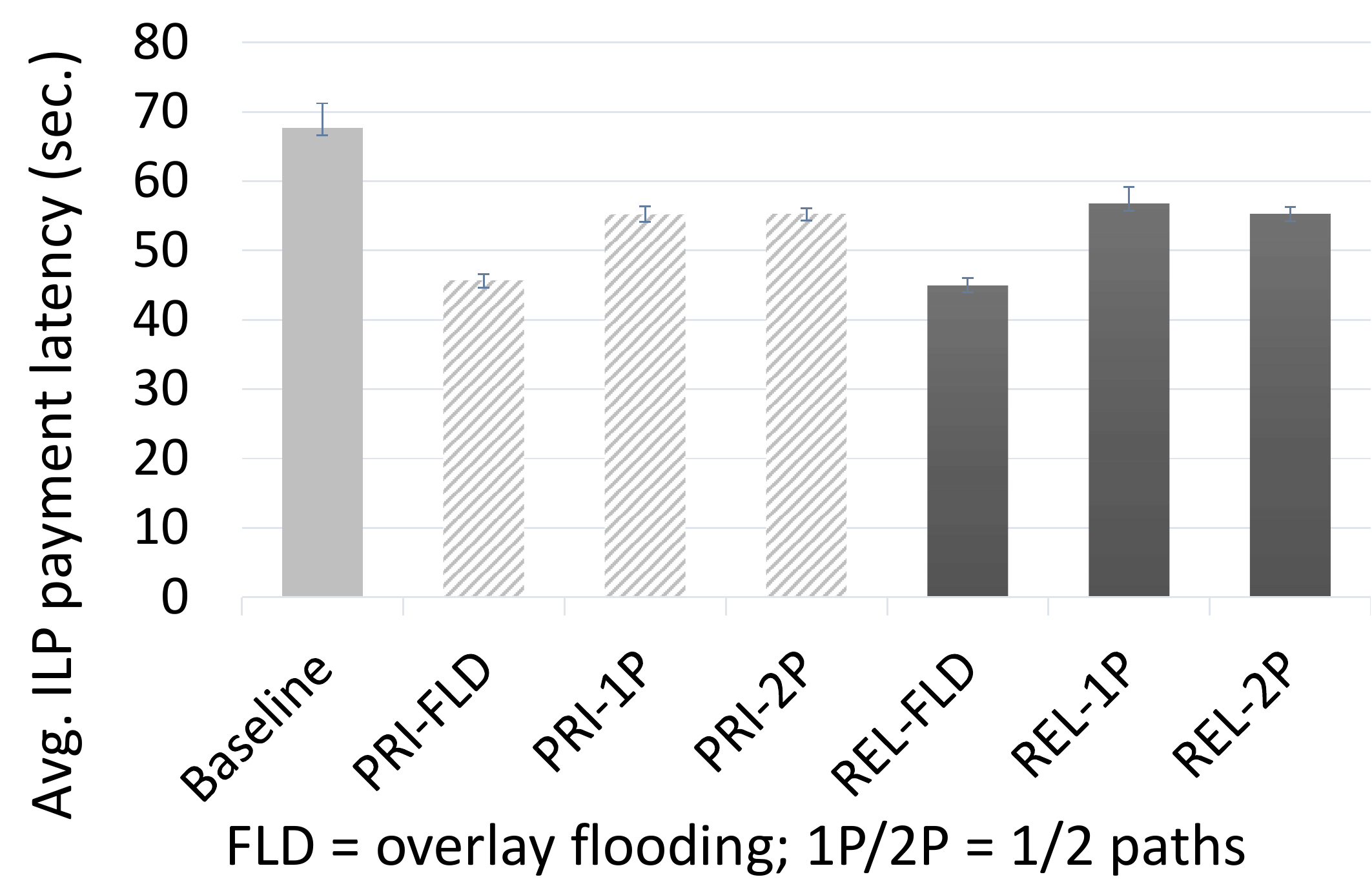}
        \caption{Loss 5\%}
        \label{fig:Exp3a_ILP_STREAM_Latency_5pct}
    \end{subfigure}
    \begin{subfigure}[t]{0.23\textwidth}
        \centering
        \includegraphics[height=1.15in]{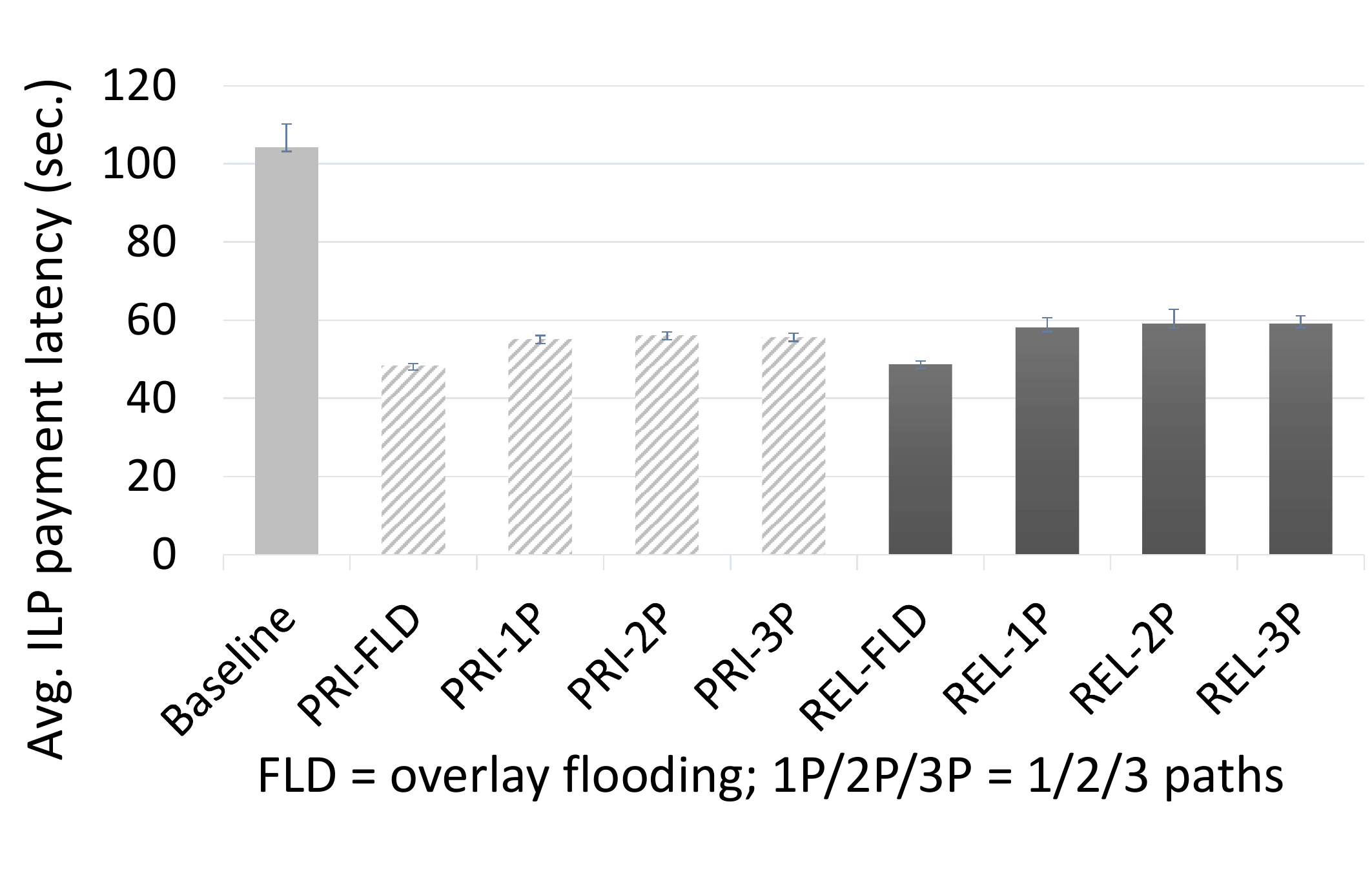}
        \caption{Loss 10\%}
        \label{fig:Exp3a_ILP_STREAM_Latency_10pct}
    \end{subfigure}
    \caption{Payment latency on the Chain topology in a network loss scenario, Priority (PRI) or Reliable (REL) messaging.}
\end{figure*}

{\em Payment latency.}
We evaluate ILP payment latency under similar scenarios with network loss.
%
On the topology in Figure~\ref{fig:chain_topology} we sent 20 ILP \textit{STREAM} payments. The amount per ILP payment was 100000 drops (1 drop = 0.000001 XRP)\footnote{https://xrpl.org/xrp.html, accessed August 2021}; each STREAM packet was 100 drops. Thus, for each payment we sent 1000 ILP STREAM micro-transactions. We used \textit{Priority} and \textit{Reliable} messaging with FLD (k=0), 1 and 2 paths (k=1,2). The loss was set again on link S12-S13. In Figures~\ref{fig:Exp3a_ILP_STREAM_Latency_0pct},\ref{fig:Exp3a_ILP_STREAM_Latency_2pct},\ref{fig:Exp3a_ILP_STREAM_Latency_5pct},\ref{fig:Exp3a_ILP_STREAM_Latency_10pct} we compare the time taken to complete the transactions over \system, with the baseline: under ideal conditions (loss 0), payment latency over SPON is a little bit larger than over the baseline (under 5\%, or 2s in this case), while at 2\% loss, \system~already offers a gain of 10\% (5s) in FLD mode. At 5\% loss, all \system~modes show 15-33\% gains.  


\subsubsection{\textit{Global topology}}
To demonstrate the behavior in a more realistic scenario, we repeat the experiments above on the Global topology; inspired from \cite{spines_dissemination_graphs}, it offers increased link redundancy while using well-chosen real-world, global locations spanning US, EU and Asia. Each circle represents an overlay node deployed on our Mininet testbed. As baseline, we sent STREAM ILP payments between two \textit{connectors} paired directly over a single link with delay 148ms - equivalent to the fastest path from Figure~\ref{fig:global}. On the global topology, the \textit{connectors} were attached to the overlay nodes FRA and HKG, and sent a total of 16 ILP payments directly through the STREAM protocol (no SPSP). The total transaction amount was 100000 drops per ILP payment, and each STREAM packet was 500 drops (200 STREAM micro-transactions). The \textit{loss} was introduced between HKG and SJC because the link belongs to multiple low latency (possible) paths, and as such, with chances to impact multiple possible flows. 

\begin{figure*}[t!]
    \centering
    \begin{subfigure}[t]{0.24\textwidth}
        \centering
        \includegraphics[height=1.15in]{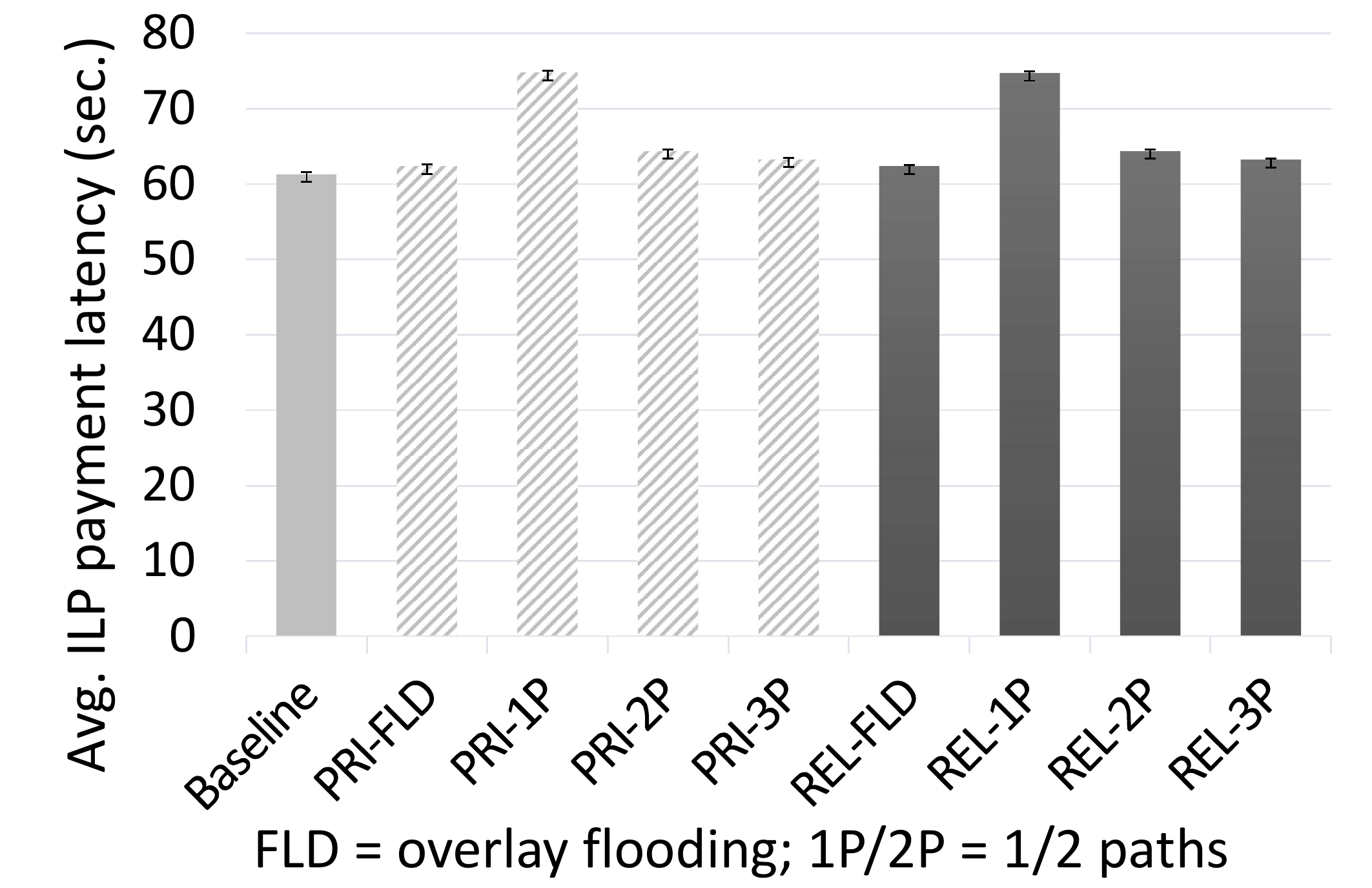}
        \caption{Loss 0\%}
        \label{fig:Loss0_Amy}
    \end{subfigure}    
    \begin{subfigure}[t]{0.24\textwidth}
        \centering
        \includegraphics[height=1.15in]{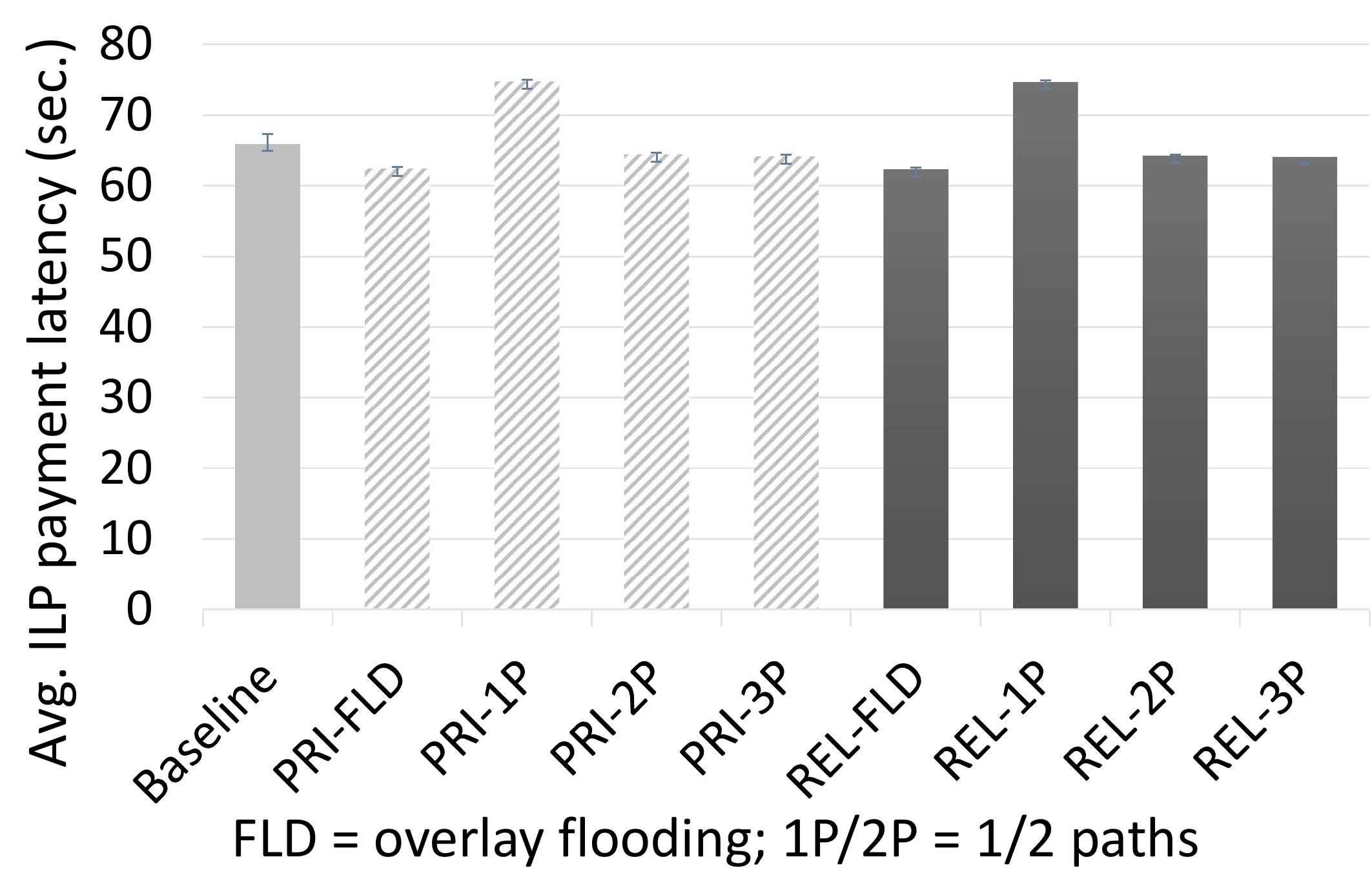}
        \caption{Loss 2\%}
        \label{fig:Loss2_Amy}
    \end{subfigure}%
    \begin{subfigure}[t]{0.24\textwidth}
        \centering
        \includegraphics[height=1.15in]{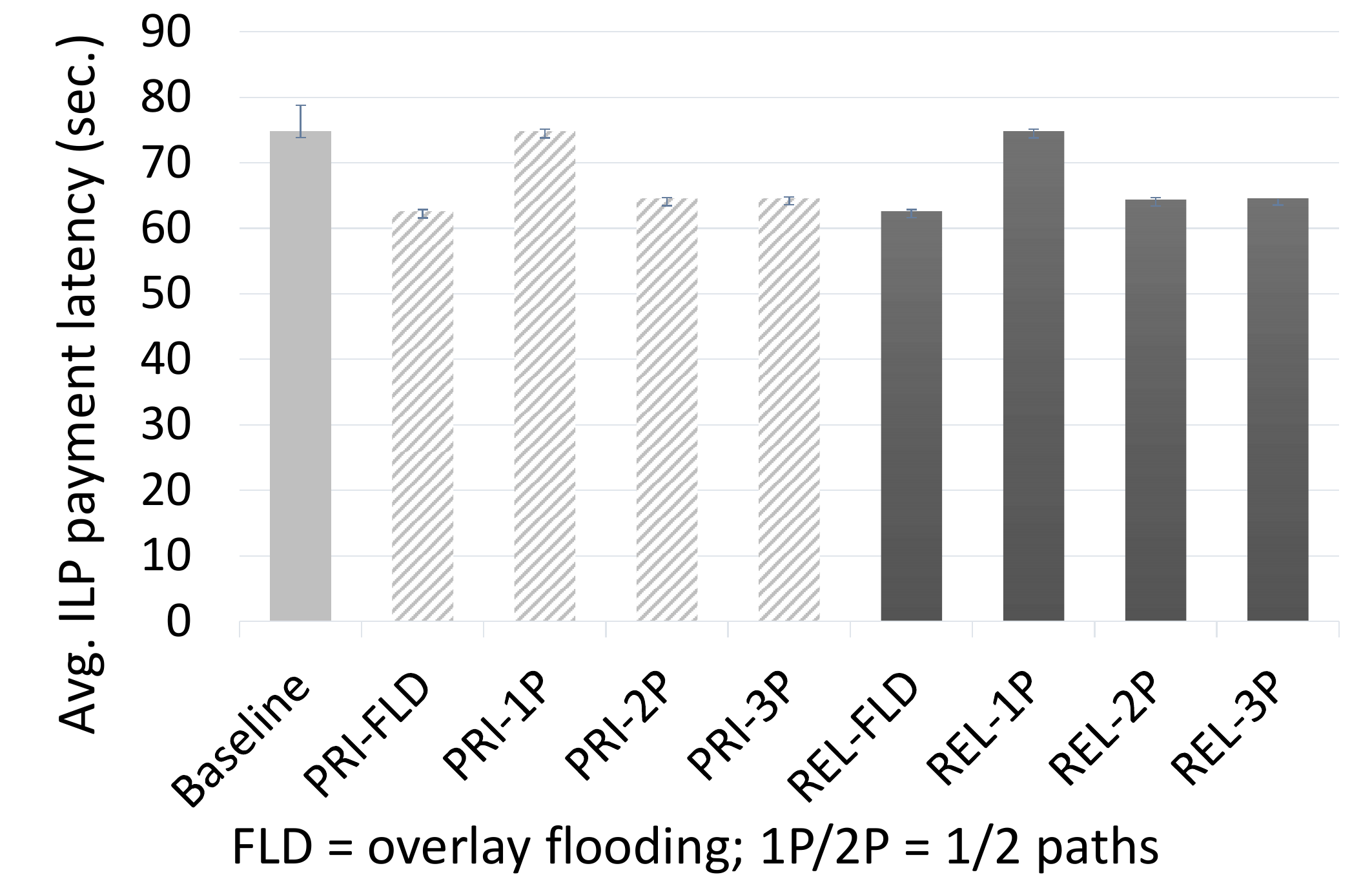}
        \caption{Loss 5\%}
        \label{fig:Loss5_Amy}
    \end{subfigure}%
    \begin{subfigure}[t]{0.24\textwidth}
        \centering
        \includegraphics[height=1.15in]{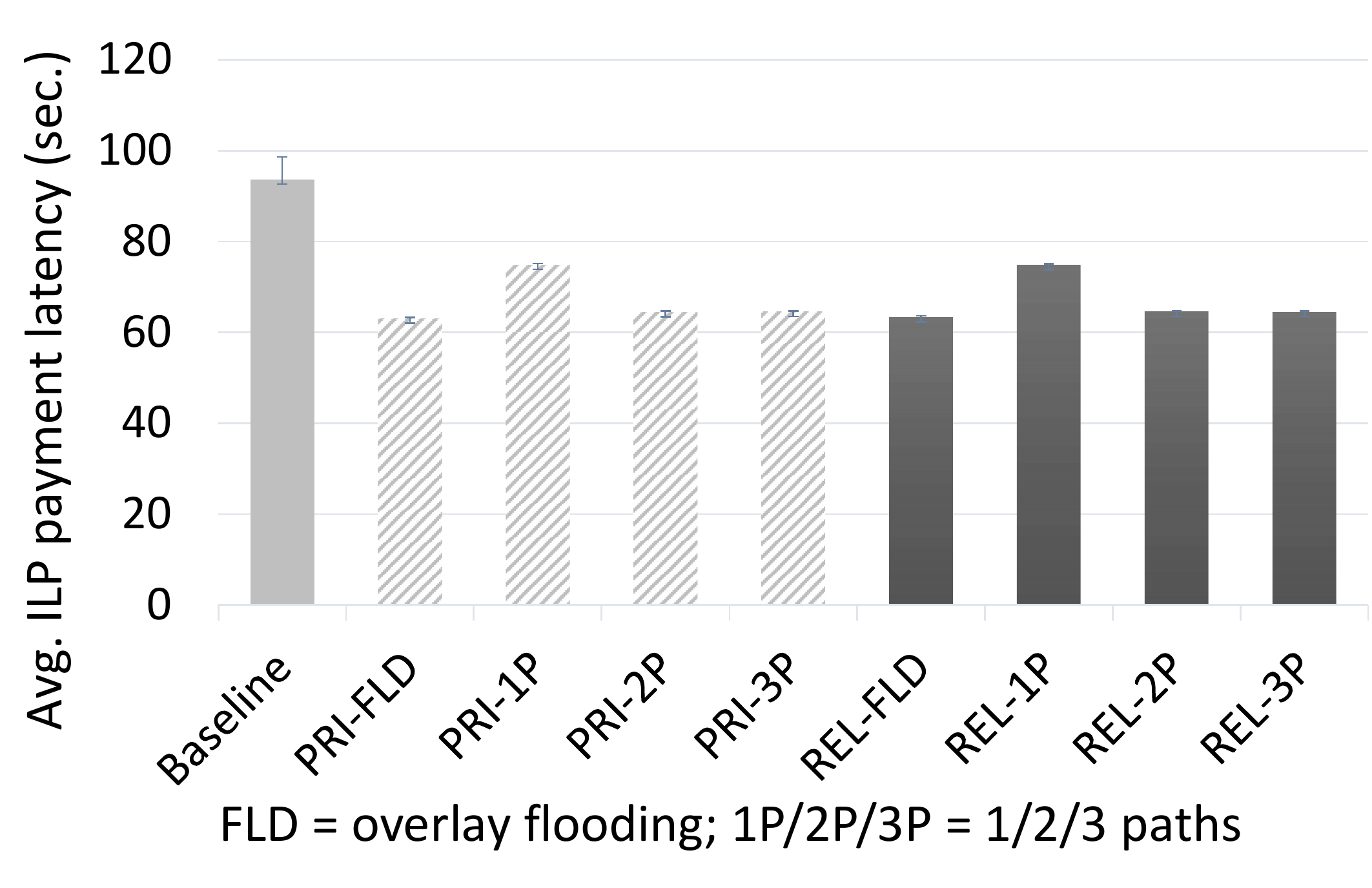}
        \caption{Loss 10\%}
        \label{fig:Loss10_Amy}
    \end{subfigure}
    \caption{Payment latency on the Global topology in a network loss scenario, Priority (PRI) or Reliable (REL) messaging.}
\end{figure*}

The results in Figure~\ref{fig:Loss0_Amy},\ref{fig:Loss2_Amy},\ref{fig:Loss5_Amy},\ref{fig:Loss10_Amy} show that in ideal conditions, except for sending on 1 path, \system~adds only 1.5\% to the total payment duration, compared to baseline; at 2\% loss, SPON offers a gain of 5\%; while at 5\%, the gain is 16\%.

In summary, in all scenarios we experimented with, the additional processing introduced by \system~and identified at loss 0 was small, and the payment system offered better performance under a link loss of 2, 5, 10\%.


\subsection{Resilience to Network Melting.}
\subsubsection{Chain topology}
We want to see how an ILP payment sent over \system~behaves when all but one path fail. We set all links to loss 0. 
Because the baseline would obviously fail in this scenario, we can only assess how \system's performance would compare with a functional baseline. So concerning baseline, we send a payment between 2 \textit{connectors} paired over a link of 20ms latency - equivalent to the remaining path 1-9-10-11-5 from Figure~\ref{fig:chain_topology}, if all other paths fail. 

\begin{figure*}[t!]
    \centering
    \begin{subfigure}[t]{0.24\textwidth}
        \centering
        \includegraphics[height=1.21in]{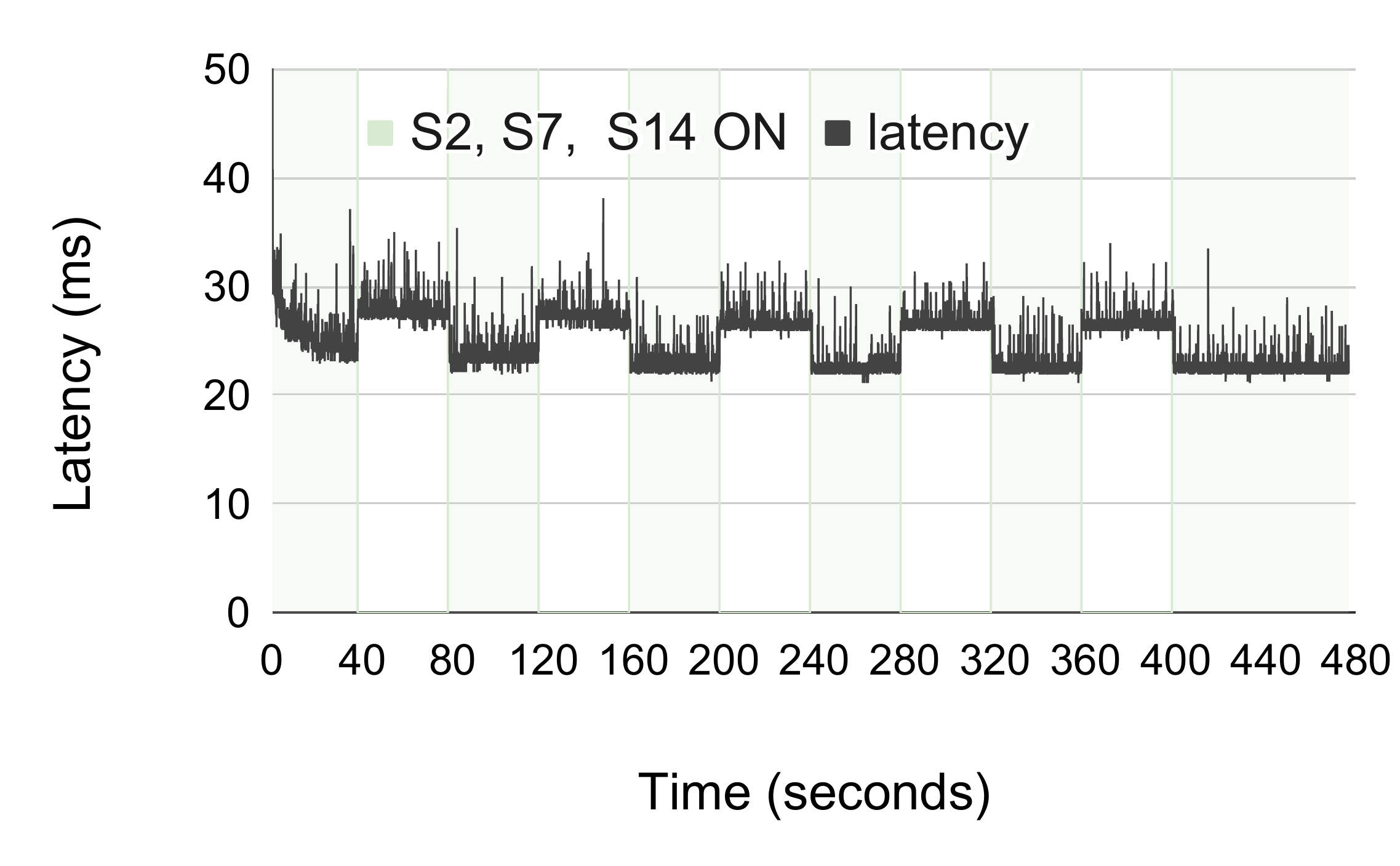}
        \caption{Flooding}
        \label{fig:ILP_latency_netw_melt_P8D1K0}
    \end{subfigure}%
    \begin{subfigure}[t]{0.24\textwidth}
        \centering
        \includegraphics[height=1.24in]{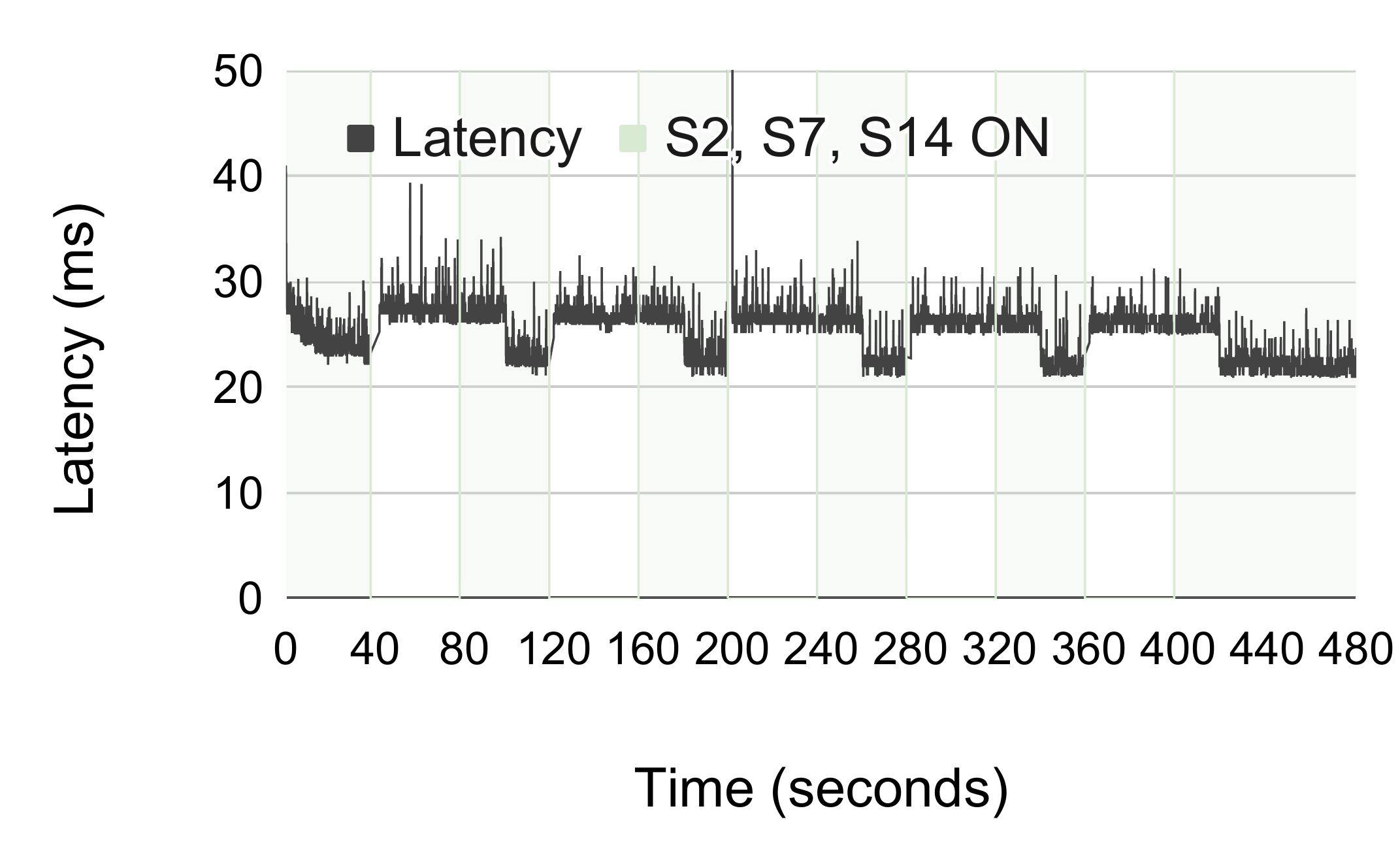}
        \caption{1-path}
        \label{fig:ILP_latency_netw_melt_P8D1K1}
    \end{subfigure}
    \begin{subfigure}[t]{0.24\textwidth}
        \centering
        \includegraphics[height=1.23in]{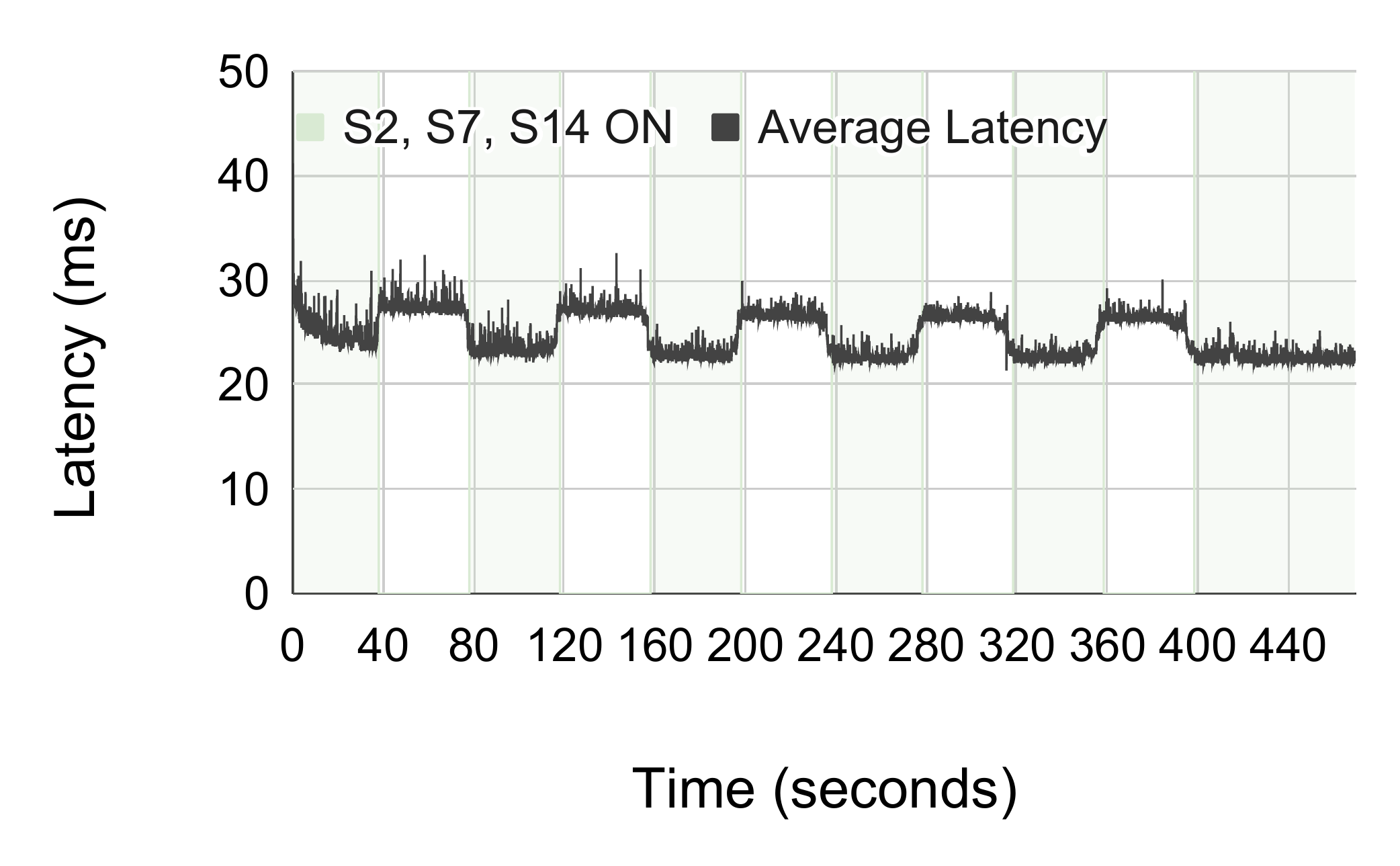}
        \caption{2-paths}
        \label{fig:ILP_latency_netw_melt_P8D1K2}
    \end{subfigure}
    \begin{subfigure}[t]{0.24\textwidth}
        \centering
        \includegraphics[height=1.15in]{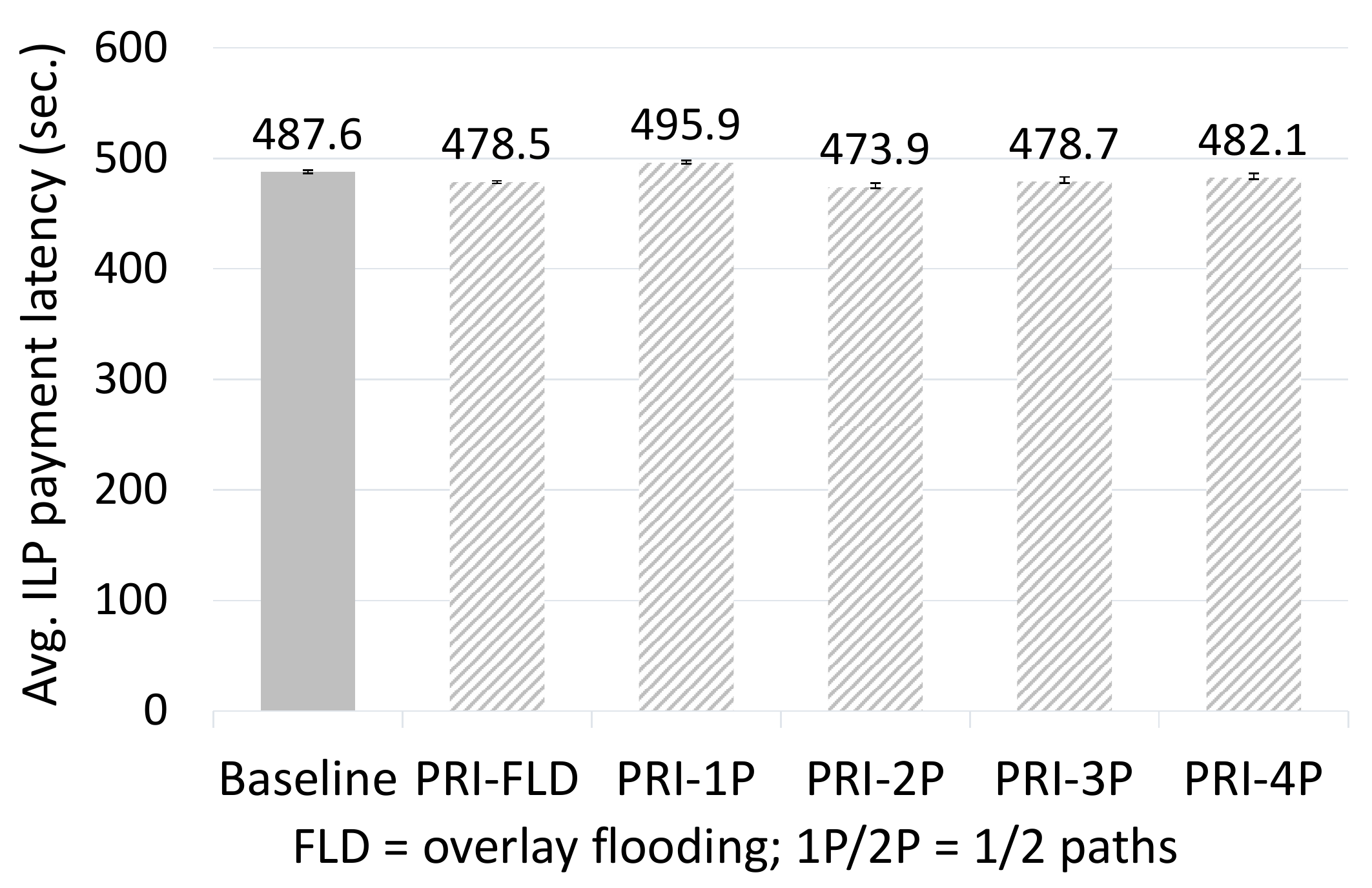}
        \caption{E2E payment latency.}
        \label{fig:paytimes_netw_melt}
    \end{subfigure}
    \caption{Payment latency on the Chain topology in a network meltdown scenario, Priority messaging (PRI).}
\end{figure*} 

We send an ILP payment of 100000 drops, and packet size 10 drops. Thus for each payment we sent 10000 ILP micro-transactions, for a total STREAM duration of 480s. While the STREAM is sent, we take down the communication of the overlay nodes {\em 2}, {\em 7}, {\em 14} using \textit{IPtables} on the respective machines, at a 40s interval, in a 5-count cycle. This procedure completely melts and brings back every 40s, all the possible paths but the green one (nodes 1-9-10-11-5) from Figure~\ref{fig:chain_topology}. 

In Figures~\ref{fig:ILP_latency_netw_melt_P8D1K0},~\ref{fig:ILP_latency_netw_melt_P8D1K1},~\ref{fig:ILP_latency_netw_melt_P8D1K2} we plot individual ILP packet latencies. We observe that, if one of the currently active transmission paths is the actual path to remain unaffected by the network melt, then the system can offer optimal protection against the melt starting even from 2-paths; on 1-path, the minimal drawback comes due to rerouting time to a better path after the network becomes available again. Even when all but one path vanish, \system~service continues reliably, with no packets lost during the experiment. \\
\indent Concerning the total duration of payments sent over the baseline versus \system, even when the latter was subjected to the severe path flipping above, it still performed slightly better than the baseline (3\%), as shown in Figure~\ref{fig:paytimes_netw_melt}. This is because the baseline is able to send only on the 20ms link, while at times, \system~can also use the fastest path of 16ms.\\
\indent{\em 2) Global topology.}
\begin{figure*}[t!]
    \centering
    \begin{subfigure}[t]{0.24\textwidth}
        \centering
        \includegraphics[height=1.2in]{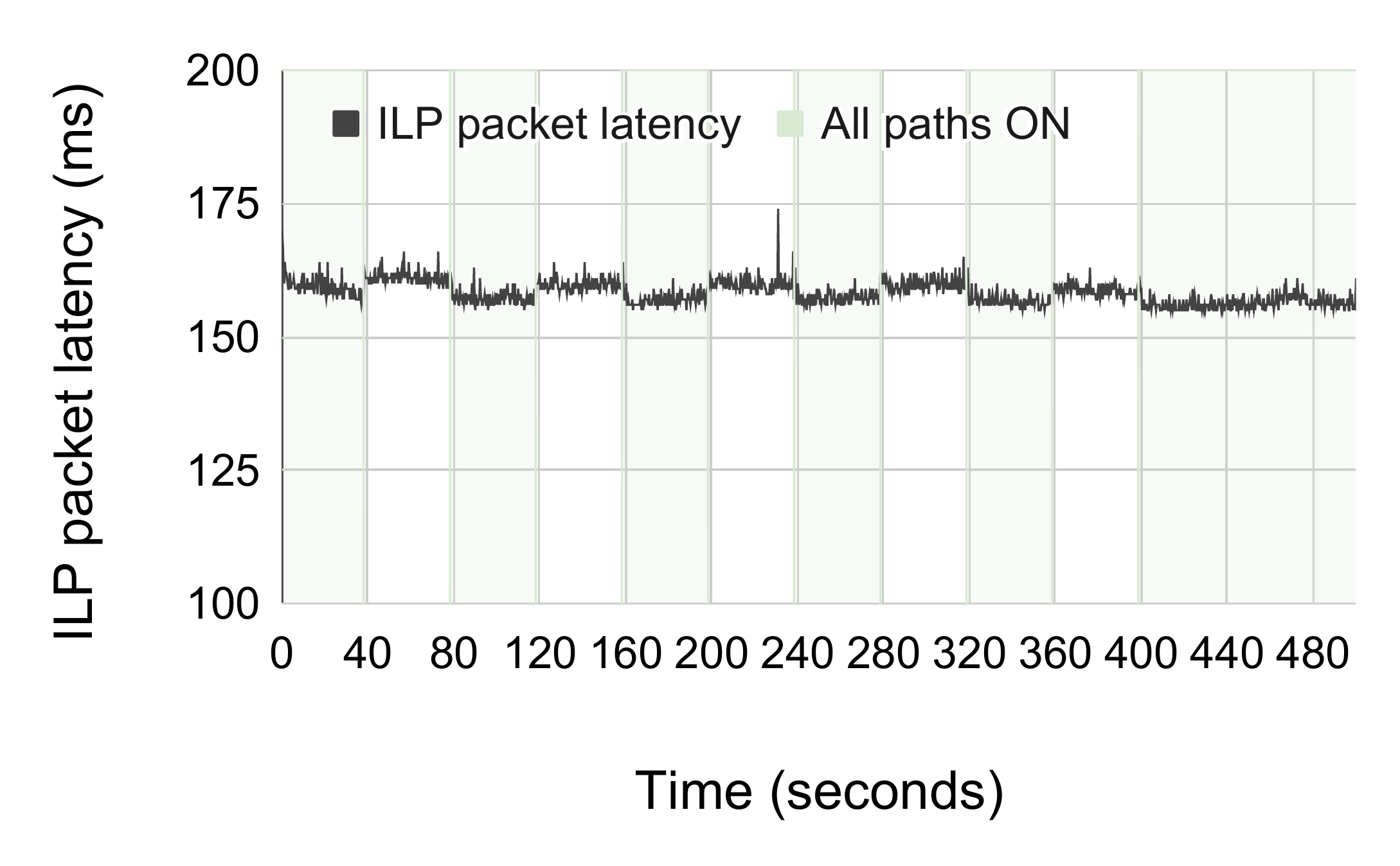}
        \caption{Flooding}
        \label{fig:Amy_melt_latency_P8D1K0-1}
    \end{subfigure}%
    \begin{subfigure}[t]{0.24\textwidth}
        \centering
        \includegraphics[height=1.2in]{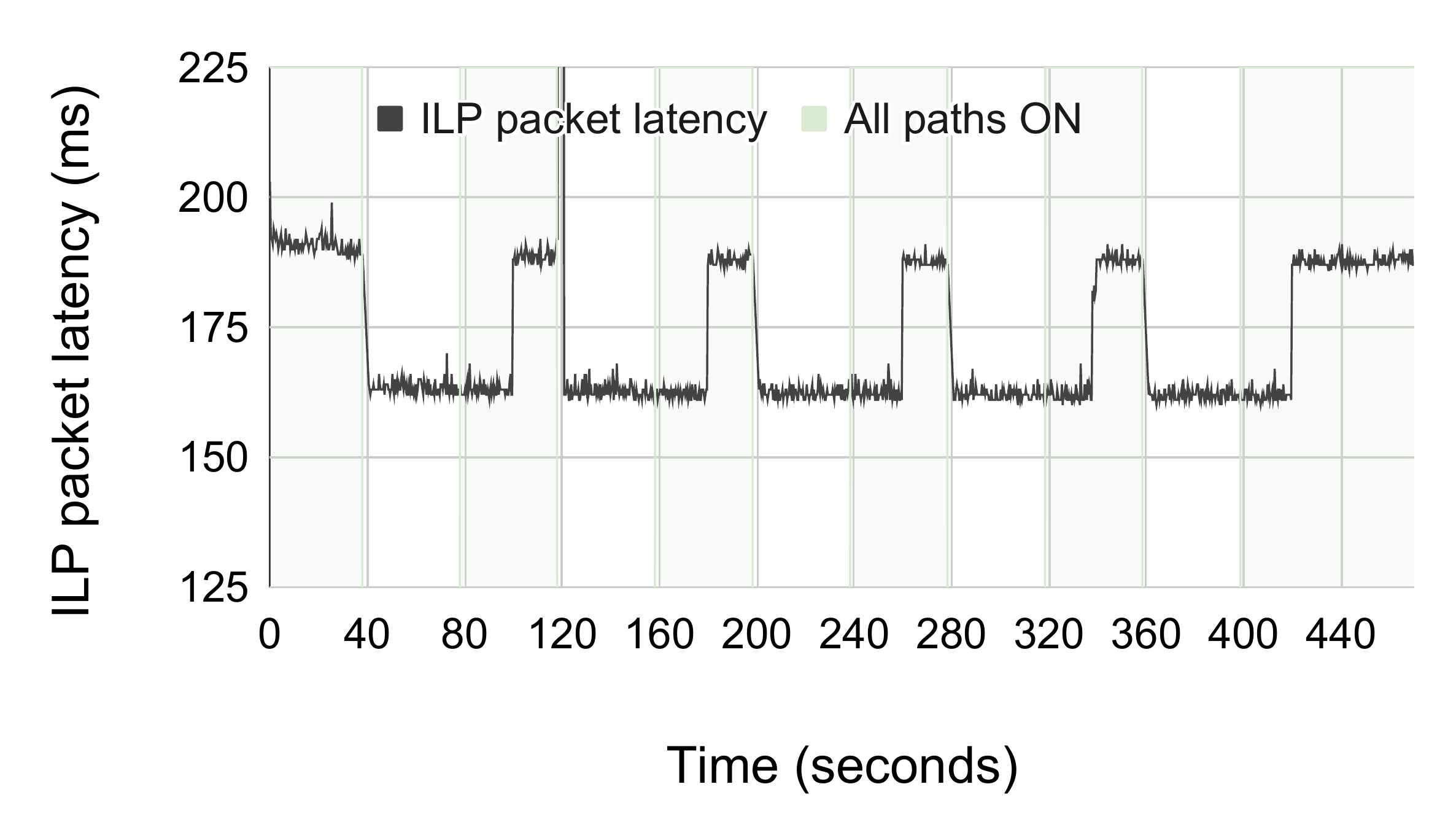}
        \caption{1-path}
        \label{fig:Amy_melt_latency_P8D1K1-1}
    \end{subfigure}%
    \begin{subfigure}[t]{0.24\textwidth}
        \centering
        \includegraphics[height=1.2in]{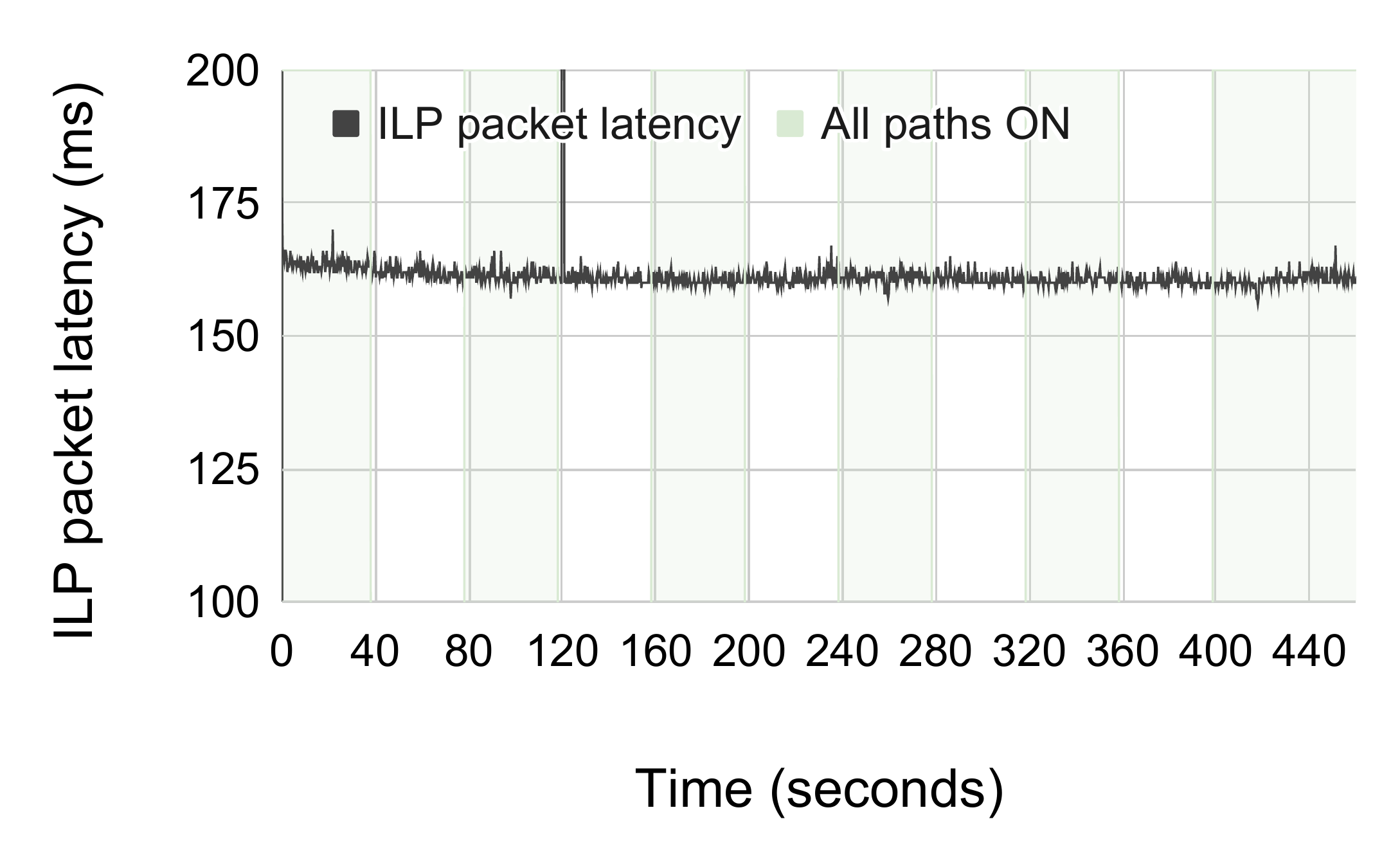}
        \caption{2-paths}
        \label{fig:Amy_melt_latency_P8D1K2-1}
    \end{subfigure}
    \begin{subfigure}[t]{0.24\textwidth}
        \centering
        \includegraphics[height=1.2in]{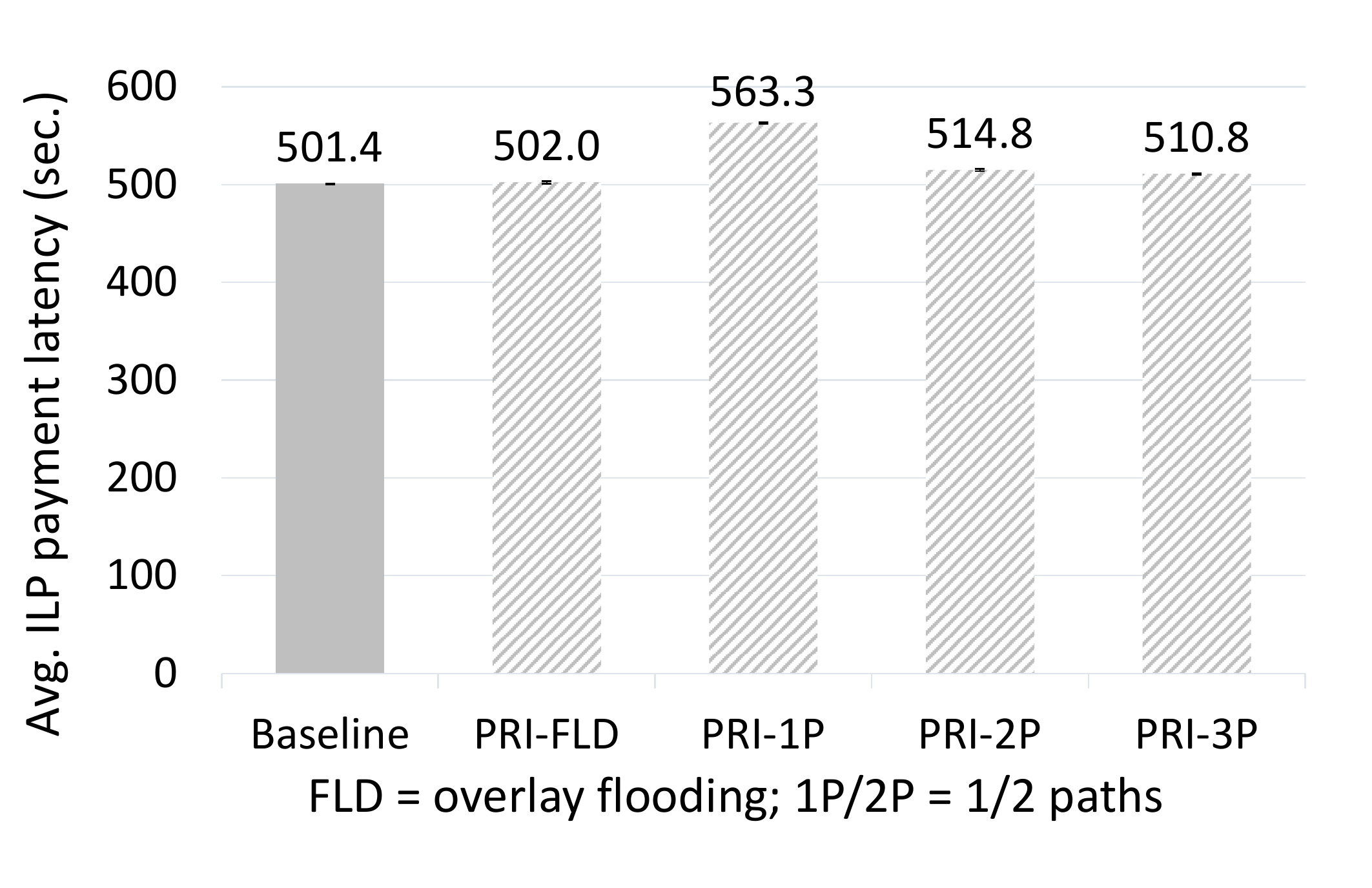}
        \caption{E2E payment latency.}
        \label{fig:Amy_melt_latency_E2E}
    \end{subfigure}
    \caption{Payment latency on the Global topology in a network meltdown scenario, Priority messaging (PRI).}
\end{figure*} 
Through our two \textit{connectors} attached to the Spines nodes FRA and HKG, we sent a payment of 80000 drops, and packet size 50 drops (1600 ILP micro-payments), during a total time of 500s. While the STREAM is sent, we cut the communication of nodes SJC, NYC, LON, WAS, JHU, DFW, ATL using \textit{IPtables} on the respective machines, at a 40s interval, in a 5-count cycle. This procedure completely melts and brings back every 40s, all the possible paths but FRA-CHI-DEN-LAX-HKG from Figure~\ref{fig:global}. The baseline is two ILP \textit{connectors} paired over a single link with delay 151ms - equivalent to the remaining path (FRA-CHI-DEN-LAX-HKG) from Figure~\ref{fig:global}, after all other paths go down. To compare the time taken to complete the transactions over the overlay versus baseline, we repeat the experiment 5 times, average the results for each case, and finally represent them in Figure~\ref{fig:Amy_melt_latency_E2E}. The individual ILP packet latencies are obtained after unique, single runs of the experiment with Priority messaging over 1, 2, 3 paths or FLD (Figures~\ref{fig:Amy_melt_latency_P8D1K0-1},~\ref{fig:Amy_melt_latency_P8D1K1-1},~\ref{fig:Amy_melt_latency_P8D1K2-1}). Results for 3-path were similar to flooding and are not illustrated. We notice that in case of a complete network melt up to 1 path, \system's service continues, while the baseline completely fails. The E2E payment latency over \system, illustrated in Figure~\ref{fig:Amy_melt_latency_E2E}, is similar to the baseline (502 vs 501s).

\subsection{Resilience to Denial of Service from Malicious Clients}
\begin{figure}[ht]
\begin{center}
    \includegraphics[width=0.485\textwidth]{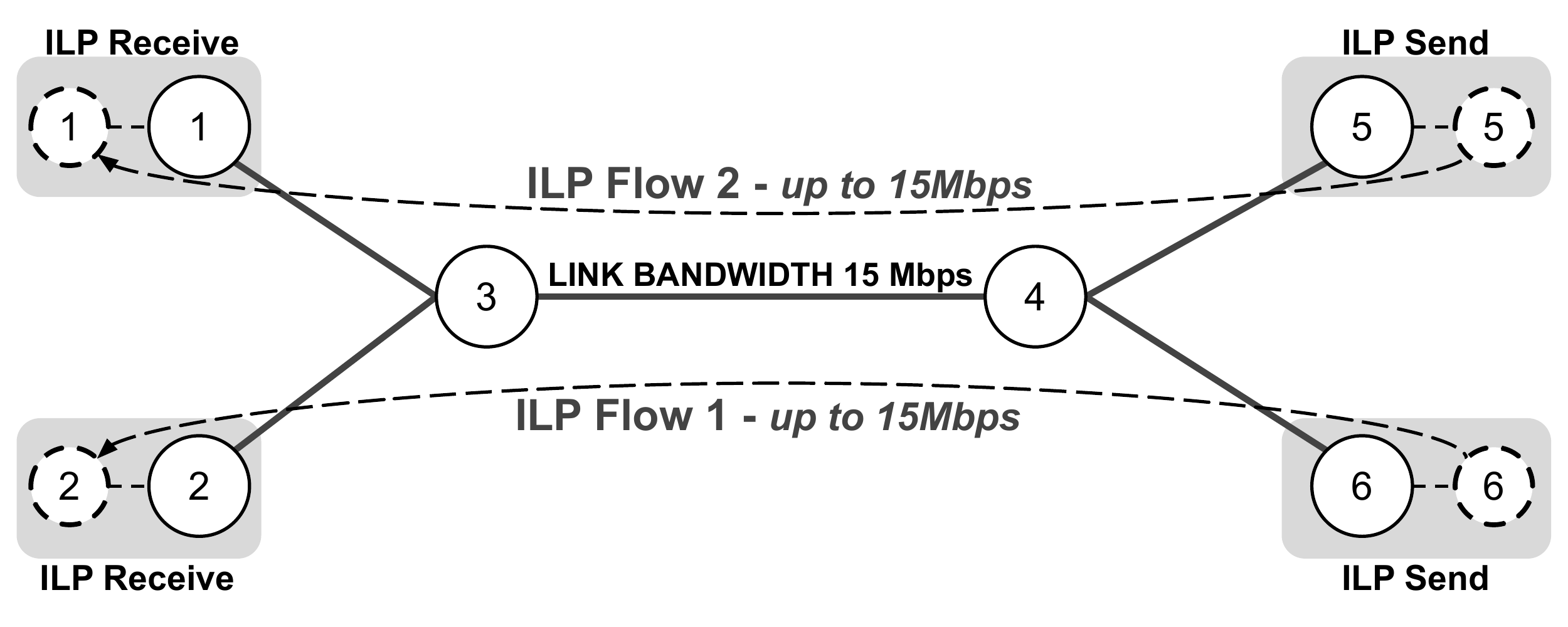}
    \caption{Network topology for the flow fairness.}
    \label{fig:Fairness_topo}
\end{center}
\end{figure}

\begin{figure}[ht]
\begin{center}
    \includegraphics[width=0.485\textwidth]{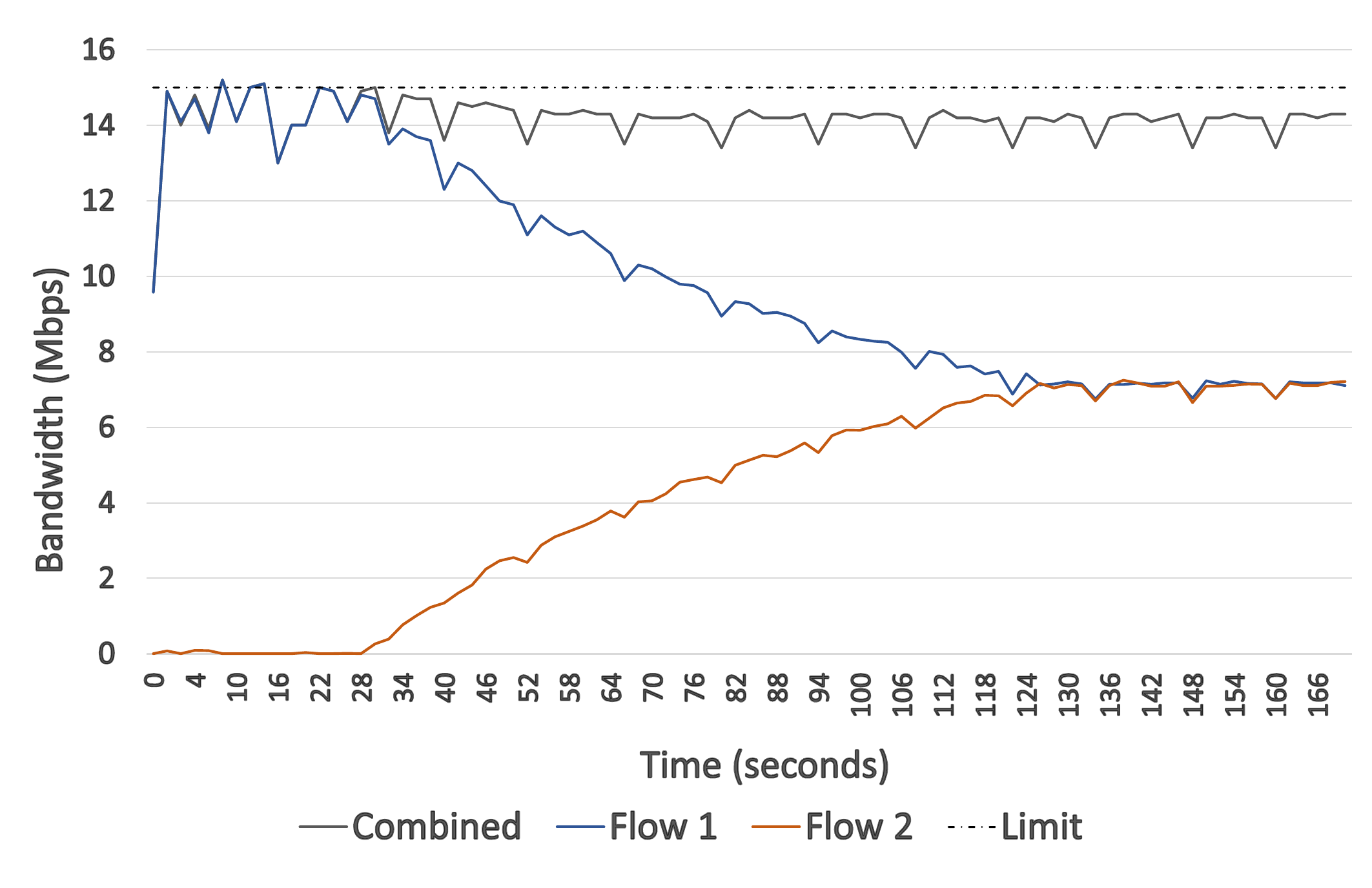}
    \caption{Legitimate and malicious flows contending for BW.}
    \label{fig:Fairness_result}
\end{center}
\end{figure}

With the aim to assess how an ILP flow sent over the overlay at maximum link capacity behaves in the presence of a second malicious flow trying to take over the channel bandwidth (BW), we attach four ILP \textit{Connectors} (1, 2, 5 and 6) to the overlay nodes 1, 2, 5 and 6 respectively (from the topology illustrated in Figure~\ref{fig:Fairness_topo}), and then we create two ILP flows. Connector 5 is paired with, and sends an "honest" flow to Connector 2 while Connector 6 is paired with, and sends a "malicious" flow to Connector 2. To each connector we can attach progressively, at 1s interval, up to 100 clients each sending over 8 streams. We are thus able to generate for each flow a maximum traffic of 15Mbps, and as such, on our topology, we set maximum link capacity to 15Mbps. For this experiment we set all links to loss 0 and as metric we used flow size in Mbps. The experiment is carried as follows. While the first, legitimate flow (C5 to C1) is sent at maximum capacity, we progressively increase the malicious, contending flow, trying to fill BW up to maximum channel capacity. Both flows were sent with Priority messaging over 1-path. As illustrated in Figure~\ref{fig:Fairness_result}, the legitimate flow decreases progressively, but only up to its fair share of 1/2 channel capacity. Although the malicious flow tried to increase its flow and send at its maximum capacity of 15Mbps, it was not able to do so beyond its fair share of BW and hence, it could not take over the channel or stop the legitimate flow. While for the particular case of ILP we experimented with only two sources, experiments with multiple sources can be found in \cite{spines_intrusion}.

\subsection{BGP Hijacking Attacks and Benign Route Misdirections.}
\begin{figure}[ht]
\begin{center}
    \centering
    \captionsetup{justification=centerfirst}
    \includegraphics[width=0.45\textwidth]{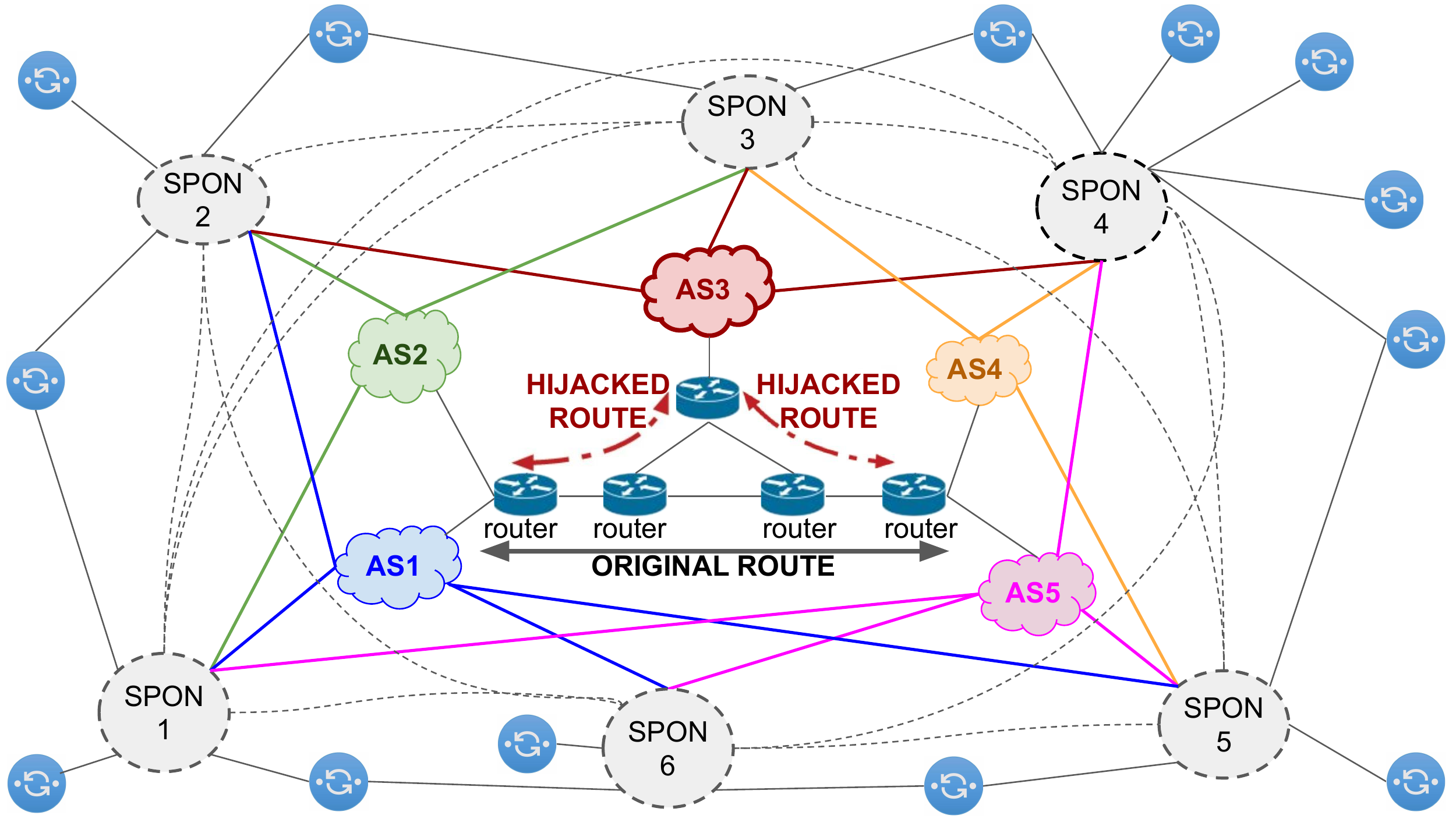}
    \caption[]{BGP attack mitigation with SPON. \\
    \smallskip
    \small{ILP nodes in light blue; overlay links between SPON nodes in dashed curvy lines; SPON connections to different ASes/ISPs in straight colored lines.}~\small\textit{Part of figure from~\cite{8613949}.}} 
    \label{fig:bgp_mitigation}
\end{center}
\end{figure}

     BGP routing attacks have been widely explored in literature. Hijacking attacks followed by double spending on Ethereum have been discussed by~\cite{8613949} for private, consortium or public deployments. An experimental topology for public networks has been illustrated in \cite{8613949}, and we use it as a working example to show how on the same topology,~\system~can defend against AS-level BGP routing attacks, through a careful design of the network. As represented in Figure~\ref{fig:bgp_mitigation}, by deploying the SPON nodes in IXPs and thus benefiting from access to say 2 or 3 ASes of interest, SPON nodes are able to ensure connectivity in spite of BGP attacks. For example, while the route between AS2 and AS4 is controlled by the adversary AS3 who partitioned AS2 from AS4, AS4 can still be reached from AS2 through SPON nodes placed appropriately in IXPs, with reduntant connections to multiple ASes, and thus still being able to relay traffic for their ILP clients located in AS2 and AS4, regardless of the hijacked route.  
   
\section{Related Work}
\label{sec:relwork}
    
    Recent efforts towards advancing the state of the art include projects like \textit{Fibre}\footnote{https://bitcoinfibre.org/, accessed May 2021}, \textit{Falcon}\footnote{https://www.falcon-net.org/about, accessed May 2021}~\cite{falconSpringer} or \textit{bloXroute}~\cite{Klarman2018bloXrouteA}, which aim to improve blockchain transaction rate by speeding up block propagation. \textit{Falcon} has the disadvantage that a block can be validated only after receiving all required packets. \textit{Fibre} uses Forward Error Correction to enable nodes to reconstruct data in advance even if some parts have been lost on the way~\cite{FIBREarxiv}, while \textit{Spines} proposes \textit{Soft Realtime Link protocol} enabling localised retransmissions to increase packet delivery ratio~\cite{FECSpines} and protects against BGP hijacking. However, all above but \system~are vulnerable to BGP failures. \textit{bloXroute} seeks to treat all blocks (or payments) fairly but it assumes that the overlay nodes can not be compromised; it also sends audit control packets (trivial to implement in SPON at ILP level using STREAM), and together with Falcon, consider the \textit{incentivization} of overlay operators (also straightforward to implement in \textit{SPON}). SABRE~\cite{DBLP:conf/ndss/ApostolakiMMV19} focuses on protecting BTC against BGP hijacking, and partially because of a low relay/client ratio, it uses software-hardware co-design to sustain high loads. It does not consider compromised relay nodes.
    Nebula~\cite{nebula} and Open Overlay~\cite{openOverlay} provide \textit{security groups} and \textit{access control lists}, but are not intrusion-tolerant. 


\section{Conclusion}
\label{sec:conclusion}

We proposed \system,~an architecture for a global payment system that uses a reliable, intrusion-tolerant overlay network. \system~provides
(1) improved payment latency, (2) fault-tolerance to benign failures such as node failures and network partitions, 
(3) resilience to routing attacks, while only incurring a small overhead. 
Our experimental results show that overlay networks are a viable solution towards making global payment systems a reality by increasing their service availability and improving latency.

\section*{Acknowledgment}
%
This work is supported by the Luxembourg National Research Fund through grant PRIDE15/10621687/SPsquared. In addition, we thankfully acknowledge the support from the RIPPLE University Blockchain Research Initiative (UBRI) for our research.  


\bibliographystyle{style/IEEEtran}
\bibliography{bib/ebpf}

\end{document}